\documentclass[fleqn,usenatbib]{mnras}
\usepackage{natbib}
\usepackage{amsmath}
\usepackage{graphicx}
\usepackage{tikz}
\usetikzlibrary{shapes,arrows}

\usepackage{listings}
\newcommand{\sqd}{sq.~deg.}

\newcommand{\mmode}{\ensuremath{m_\mathrm{mode}}}
\newcommand{\mlim}{\ensuremath{m_\mathrm{lim}}}

\title{Deep Co-Added Sky from Catalina Sky Survey Images}

\author[A Singhal et al.]{Akshat Singhal,$^{1}$\thanks{E-mail: akshats@iitb.ac.in} 
Varun Bhalerao,$^{1}$
Ashish A. Mahabal,$^{2,3}$
Kaustubh Vaghmare,$^{4,2}$
\newauthor
Santosh Jagade,$^{2}$ 
Sumeet Kulkarni,$^{5,6}$ 
Ajay Vibhute,$^{2}$ 
Ajit K. Kembhavi,$^{2}$ 
\newauthor
Andrew J. Drake,$^{3}$ S George Djorgovski,$^{3}$ Matthew J. Graham,$^{3}$ Ciro Donalek,$^{3,7}$
\newauthor
Eric Christensen,$^{8}$ Stephen Larson,$^{8}$ Edward C. Beshore$^{8}$
  \\
$^{1}$ Indian Institute of Technology Bombay, Powai, Mumbai 400076, India\\
$^{2}$ Inter-University Centre for Astronomy and Astrophysics (IUCAA), Post Bag 4, Ganeshkhind, Pune 411007, India\\
$^{3}$ Division of Physics, Mathematics, and Astronomy, California Institute of Technology, Pasadena, CA 91125, USA\\
$^{4}$ Persistent Systems Ltd., Pune, India \\
$^{5}$ Indian Institute of Science Education and Research, Homi Bhabha Road, Pune 411008, India\\
$^{6}$ Department of Physics and Astronomy, The University of Mississippi, University, MS 38677, U.S.A.\\
$^{7}$ Virtualitics, Inc., Pasadena, CA, 91101, USA\\
$^{8}$ Lunar and Planetary Laboratory,. University of Arizona, Tucson, AZ, 85721, USA\\
}
\date{Accepted XXX. Received YYY; in original form ZZZ}

\pubyear{2020}

\begin{document}
\label{firstpage}

\maketitle
\begin{abstract}
A number of synoptic sky surveys are underway or being planned. Typically they are done with small telescopes and relatively short exposure times. A search for transient or variable sources involves comparison with deeper baseline images, ideally obtained through the same telescope and camera.  With that in mind we have stacked images from the 0.68~m Schmidt telescope on Mt. Bigelow  taken over ten years as part of the Catalina Sky Survey. In order to generate deep reference images for the Catalina Real-time Transient Survey,
close to 0.8 million images over 8000 fields and covering over 27000~\sqd\ have gone into the deep stack that goes up to 3 magnitudes deeper than individual images. CRTS system does not use a filter in imaging, hence there is no standard passband in which the optical magnitude is measured. We estimate depth by comparing these wide-band unfiltered co-added images with images in the $g$-band and find that the image depth ranges from 22.0--24.2 across the sky, with a 200-image stack attaining an equivalent AB magnitude sensitivity of 22.8. 
We compared various state-of-the-art software packages for co-adding astronomical images and have used SWarp for the stacking. We describe here the details of the process adopted. This methodology may be useful in other panoramic imaging applications, and to other surveys as well. The stacked images are available through a server at  Inter-University Centre for Astronomy and Astrophysics (IUCAA).  
\end{abstract}

\section{Introduction}
Time domain astronomy is a vibrant and growing field powered mainly by synoptic sky surveys, often using relatively modest diameter telescopes (for a review of different sky surveys see e.g. \citet{djorgovski2012}). The temporal spacing in the images at a given location, along with filters, size of the mirror, total area covered etc. determine the utility of the survey for time-varying phenomena at different time-scales and depth. When looking for rare transient events, the availability of a deeper image is essential for determining the nature of the transient sources.
Ideally such comparison images will have the same sampling and filters as the survey data themselves, and this is commonly done in all synoptic sky surveys. One such survey  is the Catalina Sky Survey
\citep{Larson1998, Larson2001},
which is unique in the combination of area coverage, depth, and the number of epochs, and the overall time baseline.
All surveys, regardless of their power, will benefit from such co-addition.

The Catalina Sky Survey incorporates multiple telescopes but in this paper we focus on the 0.68~m Schmidt on Mt. Bigelow in Arizona (CSS\footnote{http://www.lpl.arizona.edu/css/}).  The original purpose of the survey has been to look for Near Earth Objects (NEOs) and it has been highly successful at it\footnote{See Figures at http://neo.jpl.nasa.gov/stats/}. 
The Catalina Real-time Transient Survey (CRTS), as the name suggests, undertakes a real-time search for transients in catalogs made from the survey data \citep[etc]{Drake2009,Djorgovski2011,Mahabal2011}. Thousands of high significance candidates have been found by CRTS and reported to the world in real-time\footnote{http://crts.caltech.edu}. The emphasis of CRTS has been to search for transients, and develop automated methods to characterize and classify them accurately despite the sparseness of available data. 
Given that spectroscopic classification is not practical for all detected transients,  even preliminary classifications based on the survey data are valuable, and deeper baseline images are a critical part of that, for example,  detecting stellar counterparts in their quiescent state or faint host galaxies of extragalactic transients. The most widely used standards reference image sets providing optical images for a large part of the sky are the Sloan Digitial Sky Survey \citep[SDSS;][]{Albareti2016} and the Pan-STARRS survey \citep[PS1\footnote{https://panstarrs.stsci.edu/};][]{2016arXiv161205243F}. 

CSS has obtained hundreds of epochs over the years covering 27000~\sqd\ and the co-added images can be deeper than SDSS or PanSTARRS. 
However, CSS images are acquired without using any filter, with the wavelength response being defined by the telescope and camera system. Hence, there is no particular band in which optical magnitude is measured. We refer to these as ``unfiltered'' images and ``unfiltered'' magnitudes, and give comparisons with some standard filters where appropriate.
On the other hand, CSS has the added advantage that the observations span several years, and it is much more likely for it to have caught multiple brightening episodes, or the occasional one in more objects (the average total effective exposure for the stacked images is 3000 seconds in CSS as against 54 seconds for the 2.5~m SDSS).
Further, the area covered by SDSS DR16 is under 15000~\sqd\  
While the larger pixels of CSS --- $2\arcsec.5$ pixels which amounts to undersampling compared to $0\arcsec.4$ of SDSS --- are a downside in crowded fields and can introduce issues in counterpart identification, the dataset is still immensely useful for various science cases involving isolated sources.

During the nightly transient detection process CRTS uses a reference median-combined stack of at least 20 images \citep{Drake2009}. A deeper stack will lead to a deeper comparison image, and allow us to rule out fainter but consistent sources when looking for transients that brighten by several magnitudes. With its large pixels, CSS co-adds are not ideal for de-blending objects close to each other on the sky, but they is still useful at higher Galactic latitudes where we do not reach the confusion limit despite reaching fainter levels by combining multiple images.

Past co-adding efforts involving astronomical surveys include Deep Sky\footnote{https://c3.lbl.gov/nugent/deepsky.html} that used 10-100 images from Palomar-QUEST \citep{Djorgovski2008}, SN Factory \citep{Aldering2002}, and Near Earth Asteroid Tracking \citep{Pravdo1999};  
Palomar Transient Factory \citep[PTF;][]{2009PASP..121.1395L}, Intermediate Palomar Transient Factory \citep[iPTF;][]{2016PASP..128k4502C}, Zwicky Transient Facility \citep[ZTF;][]{2014htu..conf...27B}, the DESI Legacy Imaging Surveys \citep{2019AJ....157..168D}, 
and a few other using SDSS and WISE images \citep{Annis2014,Lang2014,Meisner2017a,Meisner2017b}.

In this paper, we describe the co-addition of CSS images to produce deep stacks. We evaluated several co-addition software packages like Montage\footnote{http://montage.ipac.caltech.edu/}, SWarp\footnote{https://www.astromatic.net/software/swarp}, I-core\footnote{http://web.ipac.caltech.edu/staff/fmasci/home/icore.html} to see which one best suits our needs in terms of handling the image sizes and numbers as well as artifacts and sporadic World Coordinate System (WCS) issues. In \S\ref{sec:data} we describe the data, followed by comparison of various stacking methods in \S\ref{sec:methods}. We describe final implementation of the co-addition in \S\ref{sec:implementation}, and summarize and discuss future steps in \S\ref{sec:future}. 

\section{Data}\label{sec:data}

\begin{table*}
\begin{center}
\begin{tabular}{lcrrrr}
\hline
Field Type  &  Code  &  \# Fields  &  \# Fields  &  Max  &  Total images\\
&  &  &  discarded  &  images  &  included in co-adds\\
\hline
Follow--up & F & 2079 & 15 & 40 & 12139 \\
User & U & 2113 & 81 & 61 & 12312 \\
North & N & 2370 & 1 & 545 & 528880 \\
South & S & 1218 & 3 & 421 & 186703 \\
Other & -- & 11 & 3 & 12 & 62 \\
\hline
Total &  &7791 & 103 & & 740096 \\
\hline
\end{tabular}\\
\caption{\label{tab:fieldstats}CSS images taken in Follow-up, User, N and S modes. }
\end{center}
\end {table*}

To look for NEOs CSS uses a fixed sequence: four 30-second images spaced by about 10 minutes each. CSS marches from one field to another for 10 minutes, and then comes back to the first field for its next image.  These field locations are fixed within pointing errors. The field is identified with a keyword in the header.  In the data we identified 7,894 distinct field IDs. These include the North (N) and South (S) fields which are part of an all-sky grid in northern and southern celestial sphere and are used for regular observations, whereas User (U) and Follow-up (F) fields which are occasional and sporadic, defined by either a user for some off-the-grid observation or used for follow-up targets respectively (Table \ref{tab:fieldstats}). The U and F fields are typically scanned exactly thrice (in the standard 4-image mode), leading to a median of twelve visits per field. This is in stark contrast with the N and S fields, which were imaged a median of 217 times each between 2003 and 2012, the period covered here (Table \ref{tab:cssims}). Large variations are also present in number of visits for N and S fields --- for instance, parts close to the plane of the Galaxy are scanned infrequently. Adding the offset U and F fields with the N and S fields during stacking would have led to unnecessary depth non-uniformity. Hence, in this study, we have treated all four field types on par with each other, and created separate co-added stacks for each field ID.

\begin {table}
\begin{center}
\begin{tabular}{lrrrr}
\hline
 & \multicolumn{2}{c}{Total Images} & \multicolumn{2}{c}{Co-added Subset}\\
Statistic & All & NS & All & NS \\
\hline
Min &  1 &1 & 1 & 1\\
Max  &  596 & 596& 545 &545 \\
Mean &  101 & 212& 95 &199 \\
Stdev &  146 &156 & 138 &145 \\
Median &  12 &217 & 8 &205 \\
\hline 
\end{tabular}
\caption {\label{tab:cssims} Comparison of statistics of the number of epochs of all observed fields for the 2003-2012 period, and the subset selected for co-addition. This subset was created by rejecting some images as discussed in \S\ref{sec:data}. Each category has two columns: ``All'' for all fields observed in CSS, and ``NS'' for the N and S fields that make up the regular survey grid. All data are from telescope {\it 703} the 68cm Catalina telescope.
}  
\end{center}
\end {table}

A small fraction of images have incorrect WCS in their headers. 
Such images were identified and excluded from co-addition. Each image is 171\arcmin $\times$ 171\arcmin, and 4110 $\times$ 4096 pixels with a pixel size of 2\arcsec.5 $\times$ 2\arcsec.5. The uncompressed size of each image is 33~MB, and, 5-7~MB when H-compressed~\citep{White1994}, the standard way CSS stores images. The total compressed size for all images for the 2003-2012 period is about 5~TB.

\begin{figure}
\centering
\includegraphics[width=\columnwidth]{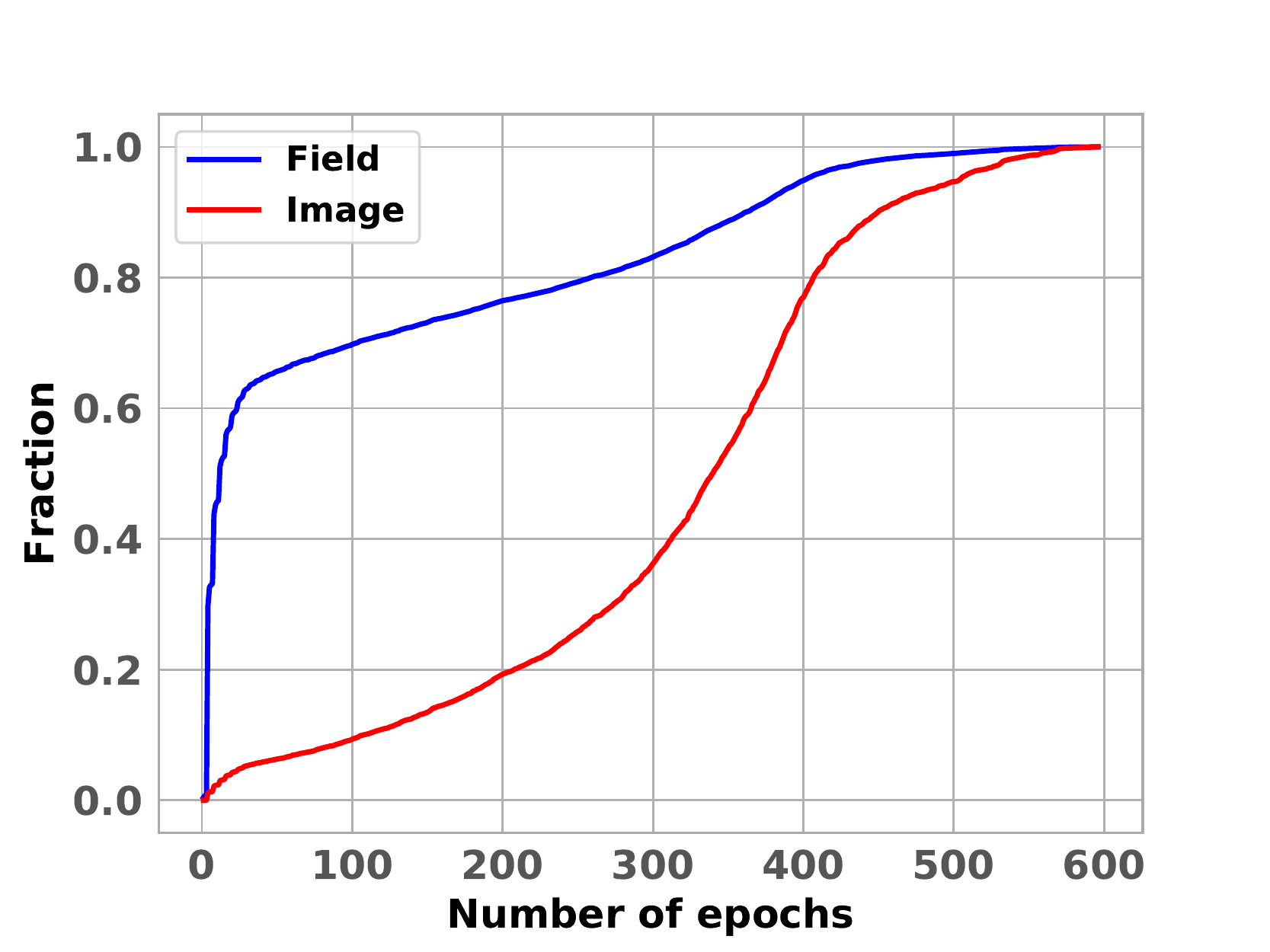}
\caption{\label{fig:fraction}The solid blue line shows the cumulative number observed fields as a function of increasing number of epochs. As defined in the text {\it number of epochs} is the number of times a field is observed. Well over half the fields were observed fewer than 30 times each, majority of which are  User  (U)  and  Follow-up  (F) fields. The solid red line indicates the cumulative fraction of total images covered by fields below a given number of epochs. }
\end{figure}

Out of 7894 fields, 4973 were observed fewer than 30 times. We use the word {\it number of epochs} (NoE) to describe how many times a field was imaged. There are 2282 fields with NoE 4, 837 fields with NoE 8, 409 fields with NoE 12 and so on. All the low NoE fields together contain only a small fraction of the total number of images (4973 fields -- 63\% of the total -- have a NoE of 30 or less, and the image contribution is $\sim5\%$ of the total. See Fig.~\ref{fig:fraction}). 


\begin{figure*}
\centering
\includegraphics[trim=0 80 0 50,clip,width=0.7\textwidth]{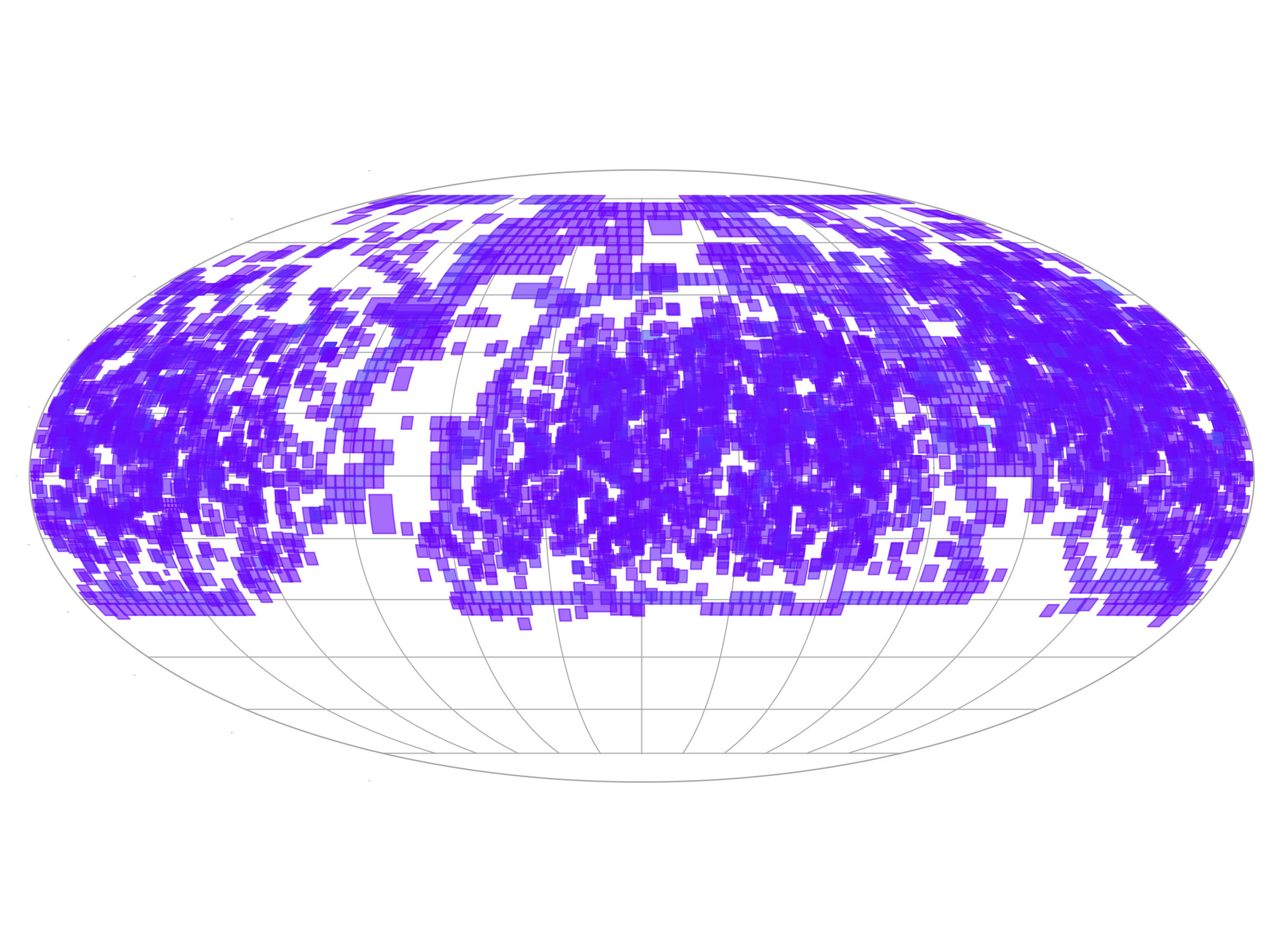}\\
\includegraphics[trim=0 50 0 50,clip,width=0.7\textwidth]{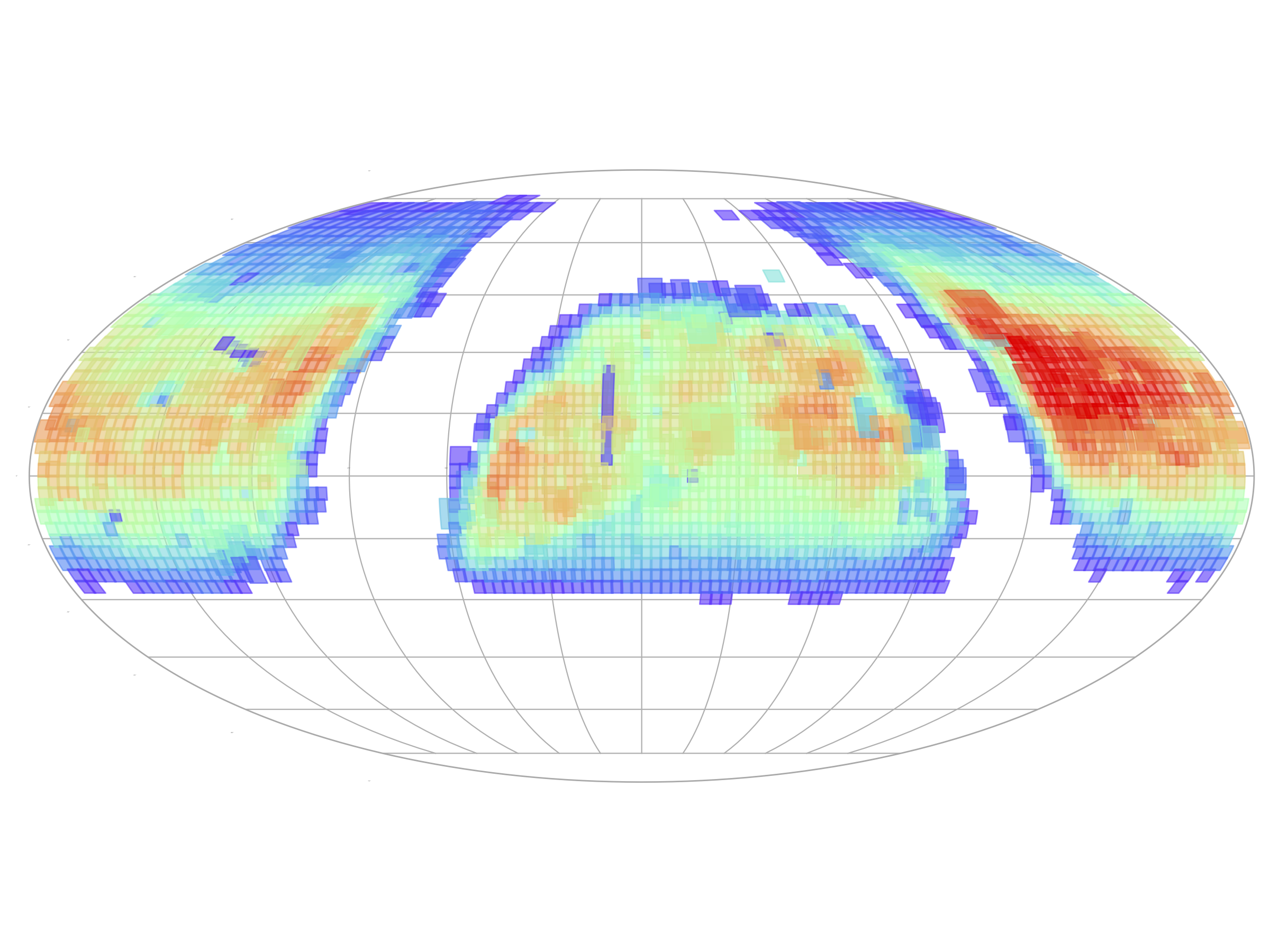}\\
\includegraphics[width=0.7\textwidth]{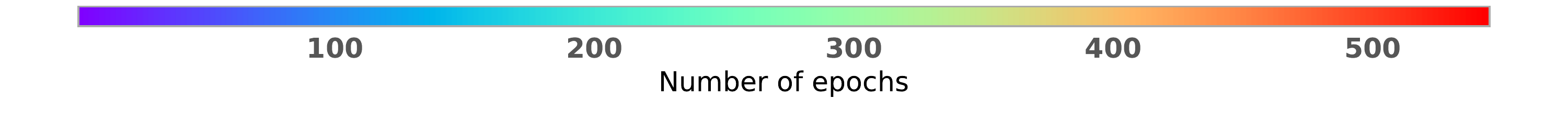}
\caption{Distribution of fields over the sky, color-coded to show the NoE (number of images): \textit{top}: shows, 4871 fields with NoE between 1-30  and \textit{bottom}: shows 2801 fields with NoE between 31-545.}
\label{fig:FreqDist}
\end{figure*}

The high NoE fields are relatively well tiled as compared to the low NoE ones which are sporadically distributed, and with differing overlap (Fig.~\ref{fig:FreqDist}).

\begin{figure}
\centering
\includegraphics[width=0.75\columnwidth]{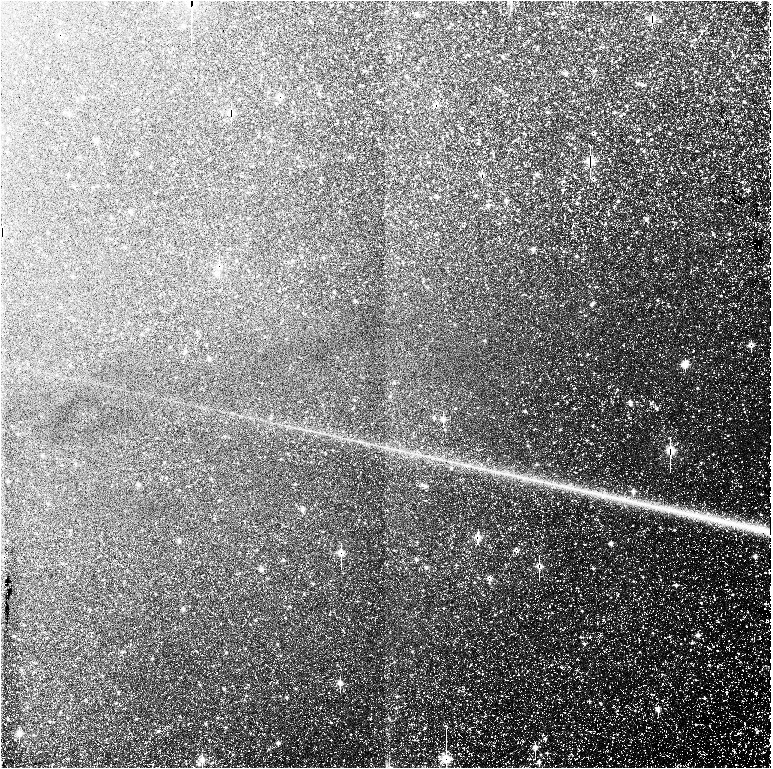}\\
\includegraphics[width=0.75\columnwidth]{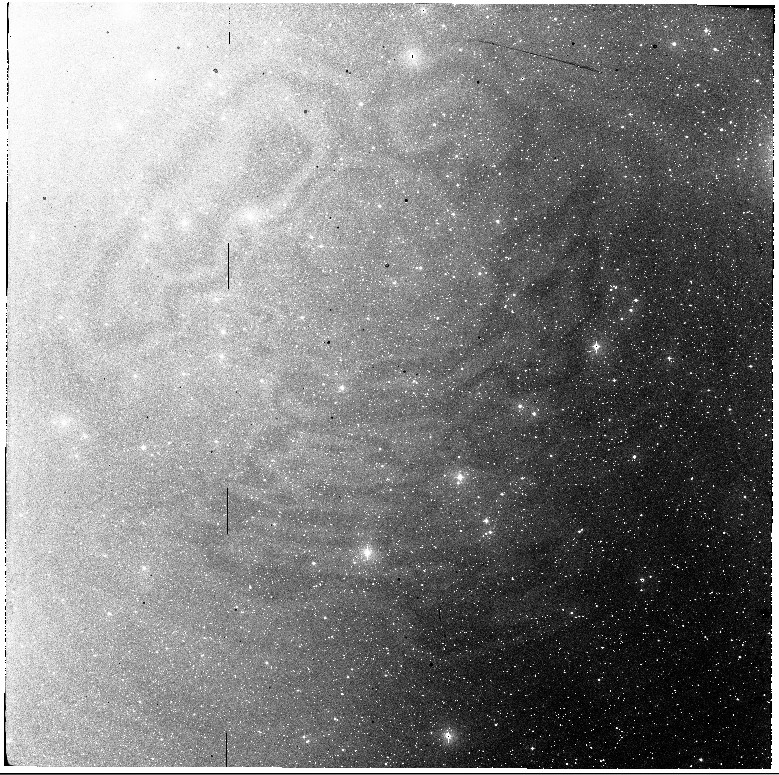}\\
\includegraphics[width=0.75\columnwidth]{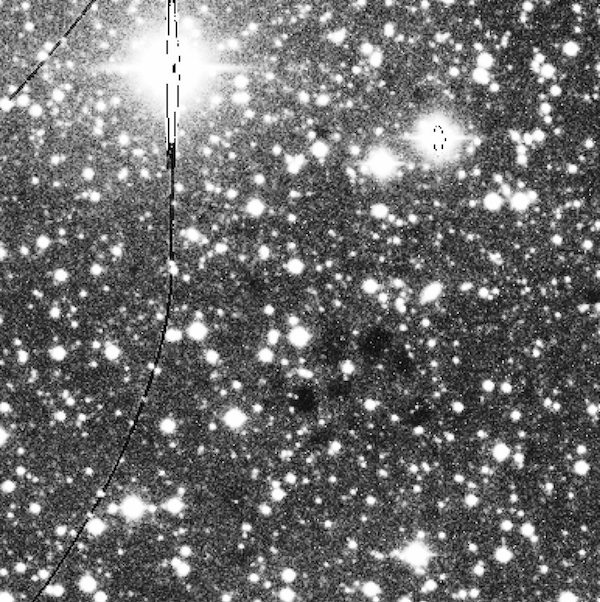}
\caption{Examples of images in CRTS repository with artifacts likely to adversely affect the quality of final co-added images. The artifacts are: \textit{Top:} A satellite trail, and discontinuous background variation across the chip. \textit{Middle:} Fringes, primarily caused by red / infrared light. \textit{Bottom:} Round holes due to dust and dirt. The knowledge of the location of these spots was part of the transient detection pipeline. They were constant over long periods, but required to be updated periodically.
Diffraction spikes and charge ``bleeding'' are seen for bright stars. 
Most of these artifacts are masked and not included in final images.
}
\label{fig:CR}
\end{figure}

CSS images show the usual artifacts present in typical imaging data, including satellite trails, CCD fringing, bleeding, spikes for bright sources, and poorer image quality for edge pixels (Fig.~\ref{fig:CR}). In particular, we note that calibrations for CSS data (e.g. flat-field images) were infrequent, which adds some uncertainty to photometry in individual images. However, we used the images that were flat-fielded during the routine runs as in our experience, reprocessing the flat fields would not have resulted in a substantial improvement in the quality of the data.

\section{Comparison of Stacking Methods}\label{sec:methods}

Many software tools, such as Montage~\citep{mont},
SWarp~\citep{Bertin1996,swarp}, and
I-core~\citep{icoreref,icoreref2}
are available for co-adding images.
We discuss some of these tools with an emphasis on Montage and
SWarp. While these two tools use different algorithms for co-addition, we found
that the final products were of comparable quality resulting in an increased depth of up to 3 magnitudes for our deepest stacks, reaching a typical sensitivity of $m_{\rm g} \sim 23$. This is roughly consistent with the central limit theorem and the depth increases as $\sqrt{N}$ increases, where N is number of co-added images.

\subsection{Montage}\label{subsec:montage}
Montage \citep{mont}\footnote{http://montage.ipac.caltech.edu/} is a WCS enabled open-source mosaicing software suite for FITS images. Montage co-adds images using ``forward pixel mapping'', where each pixel from an input image is projected onto an output image in Cartesian space such that flux and astrometric information is preserved. A direct implementation of this approach would lead to inclusion of backgrounds in the final co-adds, causing spatial non-uniformity based on exposure and also increasing the overall noise. Montage provides tools to remove the background in each image before co-adding. Montage also offers a variety of statistical methods (mean, median, weighting by number of images contributing to a pixel, etc) for combining the input images to create the output.

Rather than a single package, Montage is built as a set of tools which can be used in various combinations to create the final co-added image. We used the default statistic i.e. mean, for co-addition and compared two work flows with different methods of background estimation (Table~\ref{tab:montagetypes}). In ``MontageA'',  we use \texttt{mDiffFitExec} for comparison of each pair of input images to estimate and remove the background. Time taken to co-add the images using this method, scales as the square of the number of input images and is unwieldy for large image stacks. In ``MontageB'', the background model is created to achieve the best global fit by a least squares method. This method is faster than MontageA, as it scales nearly linearly with number of images.

The time complexity of the algorithm used by Montage goes as $\mathcal{O}(n^p)$, where $p>1$, and we can use multiple CPU cores in parallel to make co-adds faster. 
To do that we divide $N$ images into $m$ groups of $k = N/m$ images each, co-add each group separately to get $m$ temporary mosaics, which are then combined to get the final mosaic. Even the $k$ images can be subdivided beforehand if $k$ is large. Appropriate weight files are used to keep track of number of times each pixel has been co-added in the co-added image. We tested this method to a limited extent, and found that the resultant co-adds were comparable to the co-adds if done all at once.

\begin{table*}
\centering
\renewcommand{\arraystretch}{1.5}
\begin{tabular}{c|c|p{8cm}}
\hline
MontageA & MontageB & Description\\
\hline
mImgtbl & mImgtbl & Collect data from headers that comply with the FITS standard and create a table. It reports a count of images that fail that check. \\
mMakeHdr & mMakeHdr & From a list of images to be mosaicked together, generate a FITS header that best describes the output image. \\
mProjExec & mProjExec & Reproject the image \\
mDiffFitExec & $\downarrow$  &  Compare images in pairs to determine overlaps.\\
$\downarrow$  & mImgtbl & As described above\\
mBgModel & mBgModel & Determine a set of corrections to apply to each image in order to achieve the ``best'' global fit.\\
mBgExec & $\downarrow$  & Remove background from FITS images.\\
$\downarrow$  & mFlattenExec & Reject outlier pixels to fit a plane to an image, then subtract that plane from the image. \\
mAdd & mAdd & co-add the reprojected images.\\
\hline
\end{tabular}
\caption{Workflow for the MontageA and MontageB methods discussed in \S\ref{subsec:montage}. At the right we give a short description of each Montage command. More details are available at the montage web site \url{http://montage.ipac.caltech.edu/docs/index2.html}.}
\label{tab:montagetypes}
\end{table*}

\subsection{SWarp}
SWarp~\citep{Bertin1996,swarp} is part of the \texttt{astromatic.net}\footnote{http://www.astromatic.net/software/swarp} suite of packages and works with other tools like SExtractor. 
Unlike Montage, SWarp uses inverse mapping which is much faster than forward mapping of an arbitrarily large image, but arguably slightly less accurate. 
In this method, WCS information is used to obtain an outline of the image to be output, and then input pixel values are interpolated using \texttt{LANCZOS3}~\citep{swarp} method to write out the output pixels.
 The main advantage SWarp has is its robust image artifact removal~\citep{clip}. SWarp also allows image co-addition using a  variety of statistics including average, median, weighted by various parameters, sigma-clipping, etc. SWarp creates a background estimate over a grid following the SExtractor algorithm~\citep{Bertin1996} and subtracts it before co-adding the images. After various trial runs, we selected two approaches for detailed study. In the first approach (hereafter ``SwarpA''), we follow the method described in \citet{clip}, using the local standard deviation ($\sigma$) and the local median ($\mu$) to reject outlier pixels with values $>4\sigma + 0.3\mu$. In the second approach, we simply use the average of input pixels as the final output value (hereafter ``SwarpB''). Lastly, it is straightforward to provide a bad pixel mask to SWarp for ignoring certain pixels from input images. This proved to be a very useful feature, as discussed in \S\ref{subsec:compare}.

\subsection{Other methods}
In addition to Montage and SWarp, we also considered some other software tools for image co-addition. A particular case is I-core\footnote{http://web.ipac.caltech.edu/staff/fmasci/home/icore.html}, a generic FITS co-addition and mosaicing software originally written for the Wide-Field
Infrared Survey Explorer (WISE) images \citep{icoreref,icoreref2}. A key feature of I-core is to utilize small dithers between images to enhance the resolution of the final co-add. However, this requires good quality flat field images, which were not available by default for the CSS data. 

Some methods were not at a mature enough stage of development for our work. For instance, IP2\footnote{http://iccs.lbl.gov/research/isaac/IP2.html} is a software created primarily for image subtraction~\citep{ip2,ip2ref} but the team is creating an image co-addition version as well. Similarly, the Large Synoptic Survey Telescope \citep{Ivezic2015} is developing an image co-addition pipeline tuned to future LSST images. This pipeline will be configurable to other data sets, and has been tested with data from the Dark Energy Camera~\citep{Flaughter2015}. The Pan-STARRS collaboration  \citep[and references therein]{Flewlling2016} utilizes CosmoDM~\citep{Desai2015} for their image co-addition. 
\cite{2019ApJ...881L...7G} are developing a method for ZTF \citep{bellm14}.
All these software would need significant work for adapting them for the task of CSS image co-addition. As the primary focus of our research was to create deep all-sky images,  it was decided that testing all possible methods would yield a limited benefit relative to the invested effort. We also note that after the completion of our work, an optimal image co-addition technique has been published by \citet{Zackay2017a,Zackay2017b}. Here they use matched filtering and take individual PSFs into considertaion without degrading them. They apply it to simulated data and limited data from PTF \citep{Law2009,Rau2009}.

Other methods like Drizzle \citep{Fruchter2002} have been used to improve image resolution through co-addition by shrinking pixels and remapping. Given the size of the survey our main concern was quick co-addition at native scale of hundreds of images each over thousands of pointings and we did not investigate relative benefits of adopting methods like Drizzle.

\subsection{Quantitative analysis\label{sec:q}}
We arbitrarily selected 50 images each from a few fields as the test sample for comparing the co-addition methods MontageA, MontageB, SwarpA, and SwarpB. We used SExtractor to identify all sources in the image, with the ``DETECT\_THRESH'' parameter set to 3.0\footnote{Pixels that are $3\sigma$ above the local background are considered part of a star.}. We wish to compare the FWHM of sources and the depth of the co-added images across the four methods.

\begin{figure*}
\includegraphics[width=0.95\textwidth]{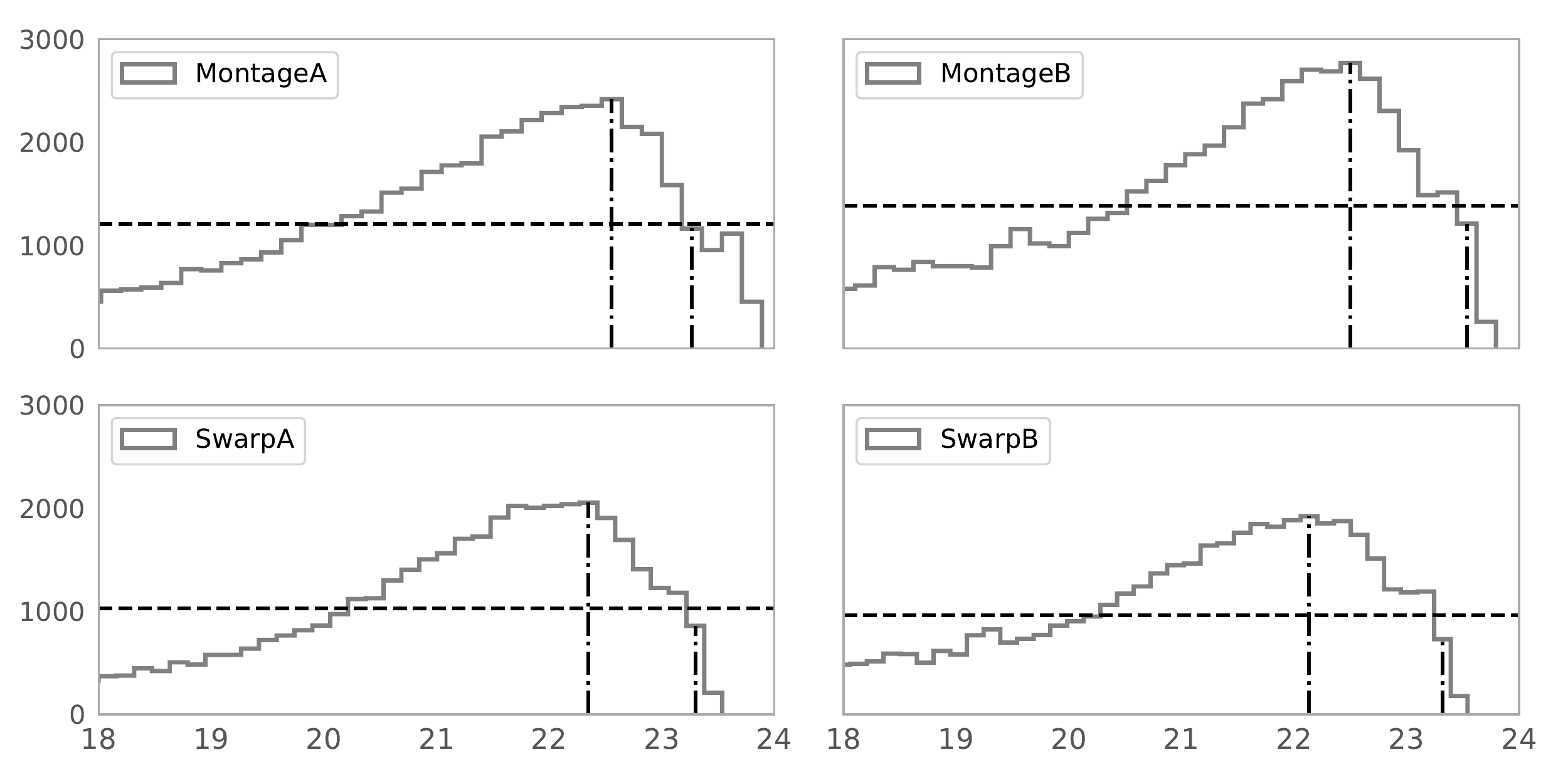}\\
\caption{Histograms of source magnitudes in the final co-added images for one set, using the four co-addition methods. 
We used SExtractor isophotal magnitudes (MAG\_ISO) and added an arbitrary but constant zero point of 28 to the instrumental magnitudes.
Vertical dot-dashed lines show the mode magnitude (\mmode) and the limiting magnitude (\mlim). The horizontal dashed lines mark half the number of sources as the histogram mode, and \mlim\ can be read off as the first bin below the rightmost intersection of this line with the histogram. 
x-axis needs label. ($V_{CSS}$)
}
\label{fig:mag_hist}
\end{figure*}

To measure depth, we need to establish a reproducible definition of the limiting magnitude. Here, we add a constant but arbitrary value of 28 to convert instrumental magnitudes to apparent magnitudes. As will be seen later, this is a reasonable value. Typical histograms of magnitudes of all sources detected in an image (Fig.~\ref{fig:mag_hist}) show increasing number of stars at fainter magnitudes upto a certain point (\mmode, the mode of the distribution), and then a decline to zero. The exact drop to zero is very likely affected by spurious sources in the image, on the other hand \mmode\ was seen to be clearly brighter than the faintest detected sources. As a result, we define the limiting magnitude \mlim\ as the magnitude bin fainter than the mode, which has half as many sources as in the mode bin. This is graphically shown in Fig.~\ref{fig:mag_hist}. Armed with this definition, we can compare depths of the co-adds created by various method by comparing the limiting magnitudes, or by the total number of sources detected in an image. The latter method is based on the reasonable assumption that bright sources will be detected in all images, and any differences in source counts are likely caused by the presence of fainter sources.

\begin {table}
\begin{center}
\begin{tabular}{lccccc}
\hline
Method&  Mode  &  Mag  &  FWHM  &  Sources & Time\\
& \mmode & \mlim & pixels & ($3.0\sigma$) & (s)\\
\hline
MontageA & 22.5 & 23.3  & 3.6& 52519 &4079\\
MontageB & 22.5 & 23.5 & 3.4& 58498 &2528\\
SwarpA & 22.3 & 23.3 &2.9& 44648 &598\\
SwarpB & 22.1 & 23.3 &3.2& 50422 &262\\
\hline
\end{tabular}
\caption{\label{tab:sextr} Comparison of the results of 50 images added together, using four different algorithms, two each from Montage and SWarp. An arbitrary but constant zero-point of 28 was used. \mlim is the limiting magnitude. The FWHM (Full Width at Half Maximum) reported in pixels is the average for all stars. The last column shows the time taken by the co-addition to complete. 
} 
\end{center}
\end {table}

We find that in the four methods compared, MontageB consistently has the faintest \mlim, while SwarpB has the poorest. It is non-trivial to account for spurious sources detected in the image, which may include image artifacts (diffraction spikes, unmasked bad pixels, cosmic rays, etc) and random noise. In this context, it is not surprising that both Montage methods lead to more sources in the final co-added image as they do not reject any pixels. This is thus a trade-off between using all data, potentially boosting the signal-to-noise ratio for brighter sources, and retaining all noise, degrading data quality at the faint end. 

To better understand the variation in total number of sources, we cross-matched the source catalogs created by each of the four methods using a matching radius of 2\arcsec\ (Table~\ref{tab:commonfrac}). We find that SwarpA is most effective in finding sources detected by other methods. In other words, given a source catalog created by any algorithm, SwarpA has the highest overall reproducibility for finding sources in that catalog. This, along with the fact that SwarpA co-adds seem to have the least number of sources in them, suggests that SwarpA is more effective at suppressing spurious sources.

\begin{table}
\centering
\begin{tabular}{lcccc}
\hline
Method $\longrightarrow$ & MontageA & MontageB & SwarpA & SwarpB\\
Comparison $\downarrow$ & & & & \\
 \hline
MontageA & 1.00 & 0.86 & 0.70  & 0.70  \\
MontageB & 0.77 & 1.00 & 0.65 & 0.66  \\  
SwarpA & 0.82 & 0.85 & 1.00 & 0.83  \\
SwarpB & 0.73 & 0.77 & 0.73 & 1.00 \\
\hline
\end{tabular}
\caption{Fraction of sextractor detected sources common to various methods. The numbers in each column denote the fraction of sources for that method detected in co-adds created by other methods. For instance, 86\% of sources in the MontageB co-add were also detected in the MontageA co-add. }
\label{tab:commonfrac}
\end{table}

CRTS images are unfiltered, which makes it difficult to compare the measured magnitudes from co-added images against a standard database for examining the quality of photometry. However, we can undertake pairwise comparison of photometry from co-adds created by each method. For instance, we cross-match the photometric catalogs from MontageA and MontageB, and for each common source we calculate the difference in magnitudes obtained from both co-adds. We find that for these two methods, the difference is 0.02~mag with a standard deviation of 0.24~mag. Similar comparisons for all methods are given in the top part of Table~\ref{table:pair}. We see that the scatter in magnitudes is higher for the brightest and faintest sources. As a result, we refine the comparison by plotting magnitudes measured from two methods against each other, and using a subset corresponding to a linear region of the plot. Values for these subsets are given in the lower part of Table~\ref{table:pair}.

\begin{table}
\centering
\begin{tabular}{l|c|c|c}
\hline
 & MontageA & MontageB & SwarpA \\
 \hline
\multicolumn{4}{c}{$ \textbf{All source}$}\\
\hline
MontageB & 0.02 $\pm$ 0.24   &
&  \\
SwarpA & $-0.01 \pm$ 0.22& $-0.02 \pm$ 0.23& \\
SwarpB & $-0.06 \pm$ 0.24  & $-0.07 \pm$ 0.21 &-0.05 $\pm$ 0.22  \\
\hline
\multicolumn{4}{c}{$ \textbf{Subset}$}\\
\hline
 MontageB & 0.01 $\pm$ 0.25  &&   \\
 SwarpA &$-0.01 \pm$ 0.15 & 0.01 $\pm$ 0.12 & \\
SwarpB &  $-0.05\pm$ 0.20  & $-0.04 \pm$ 0.15& -0.06 $\pm$ 0.12  \\
\hline
\end{tabular}
\caption{\label{table:pair}Pairwise difference in magnitude for all the common sources (``All source'') and for only those sources which follow a linear relation (``Subset''). Ideally difference in magnitude of all common sources should have been constant, but that was not the case. }

\end{table}

\begin{figure}
\includegraphics[width=\columnwidth]{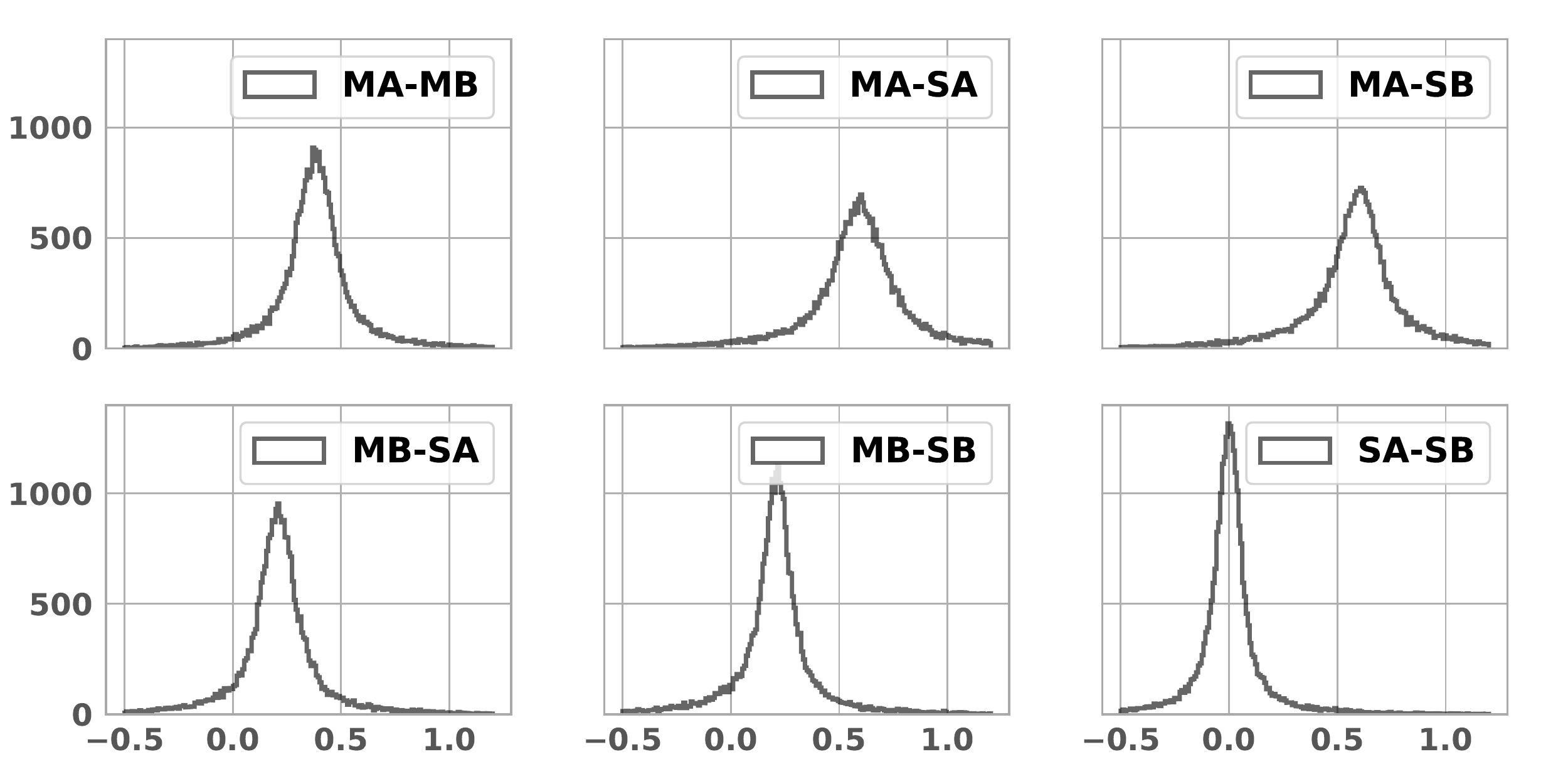}
\caption{Distribution of pairwise FWHM difference. X-axis represents $\Delta FWHM$ in pixels and Y-axis represents count. The difference in FWHM is in the same order as in the legend. E.g. the first figure in the top row represents pixel-FWHM of MontageA - MontageB. The mean and standard deviation values are given in Table~\ref{tab:pairfwhm}.}
\label{fig:diff_fwhm} 
\end{figure}

Lastly, we compared the FWHM of sources in the final co-adds by different methods. To eliminate spurious sources and a few genuine extended sources, we restricted this comparison only to sources with FWHM $<$ 20~pixels. We find that SwarpA produces sharper images of stars as compared to the other three methods (Table~\ref{tab:sextr}). This difference is seen more clearly by pairwise comparisons of the FWHM analogous to the magnitude comparisons above. Fig.~\ref{fig:diff_fwhm} shows histograms of the differences in FWHM for all objects, showcasing the sharper PSF obtained by SwarpA.  However, it may seem contradictory with the fact that Table~\ref{tab:sextr} shows a difference of 0.3 in average FWHM but considering  that it shows average FWHM of all the sources including the artifacts and Table \ref{table:pair} represents the difference in FWHM among only the common source, the discrepancy  vanishes.  

In the end we conclude that MontageA and MontageB produce slightly deeper co-adds, while SwarpA produces sharper co-adds. The key difference was that Swarp methods were modular and fast. Although we have discussed the possibility of using multiple cores in parallel for single co-adds to reduce the time, it was found that it is far more efficient to co-add one field per core.

\begin {table}
 \label{tab:psf} 
\begin{center}
\begin{tabular}{lccccc}
\hline
  &MontageA & MontageB & SwarpA &SwarpB\\
  \hline
Set 1 &  $3.7 \pm 1.8$ & $3.4 \pm 1.7$ & $3.0 \pm 1.4$ & $3.3 \pm 2.2$ \\
Set 2 &  $4.0 \pm 2.2$ & $4.1 \pm 2.6$ & $3.5 \pm 1.8$ & $3.7 \pm 2.5$  \\
Set 3 &  $3.5 \pm 1.8$ & $3.2 \pm 1.9$ & $2.8 \pm 1.4$ & $3.0 \pm 1.8$  \\
\hline
\end{tabular}
\end{center}
\caption{Mean and standard deviation of FWHM (in pixels) of common sources detected in co-added images using different methods for a few image sets. As the standard deviations are comparable to the mean, we calculate pairwise differences for the methods and report them in Table~\ref{tab:pairfwhm}.}
\end {table}

\begin {table}
\label{tab:psf2} 
\begin{center}
\begin{tabular}{lccccc}
& MontageB & MontageA & SwarpB\\
 \hline
 SwarpA & +0.21 $\pm$ 0.23& +0.58 $\pm$ 0.27  &+0.01 $\pm$ 0.21 \\
 SwarpB & +0.20 $\pm$ 0.23 & +0.58 $\pm$ 0.27  &  \\
  MontageA & -0.38 $\pm$ 0.21 & & \\
\hline
\end{tabular}
\end{center}
\caption{\label{tab:pairfwhm} Pairwise difference of FWHM (in pixels) between different methods, for sources which are common in all methods. Each value gives the difference between the column name and the row name: for instance, the FWHM of sources in MondageB is $0.21 \pm 0.23$ pixels larger than FWHM measured in SwarpA coadds. The distributions of such differences are plotted in Fig.~\ref{fig:diff_fwhm}.}
\end {table}

\begin{figure}
\includegraphics[width=\columnwidth]{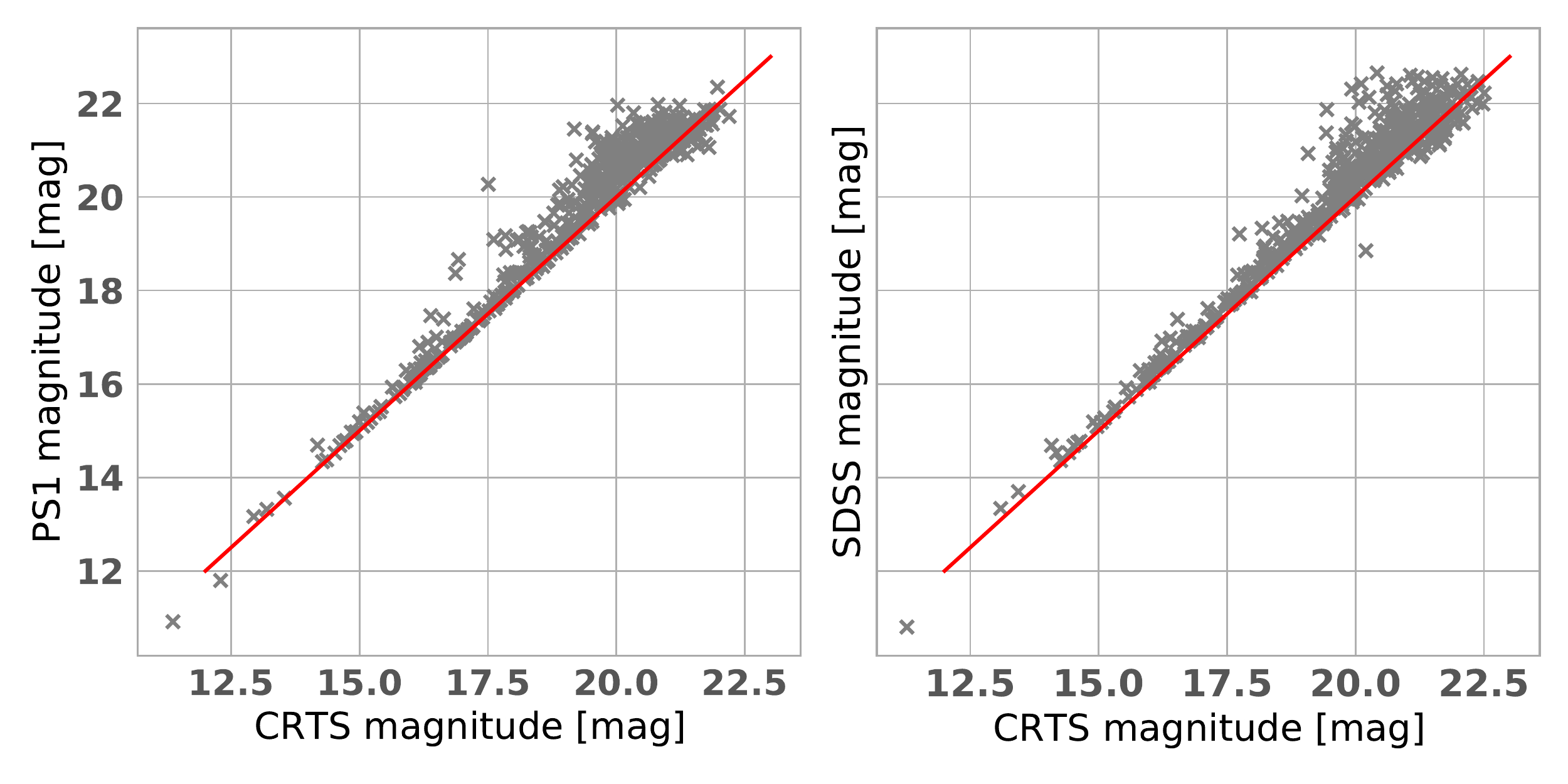}
\caption{\textcolor{black}{Constant zero point offset added to the SExtractor MAG\_ISO of the CRTS sources such that the magnitude of the SDSS and PS1 (left to right) catalog roughly match the common sources. The scatter plot shows comparison of SDSS and PS1 magnitude with offset SExtractor magnitude superimposed with slope 1 linear line. } }
\label{fig:survey_zero} 
\end{figure}

\subsection{Comparison and method selection}\label{subsec:compare}
Three of the four methods, which used direct averaging for creating the final co-add, give similar results. However, co-adds produced by the SwarpA clipped mean algorithm are significantly more robust to the presence of  artifacts present in individual images (Fig.~\ref{fig:CR}). Due to small dithers between exposures of the same field, the border regions of co-adds typically have fewer exposures. This reduces the effectiveness of the clipping procedure. However, background subtraction does a reasonable job of cleaning up these image areas.

\begin{figure}
\centering
\begin{tabular}{c}
\fbox{\includegraphics[width=60mm]{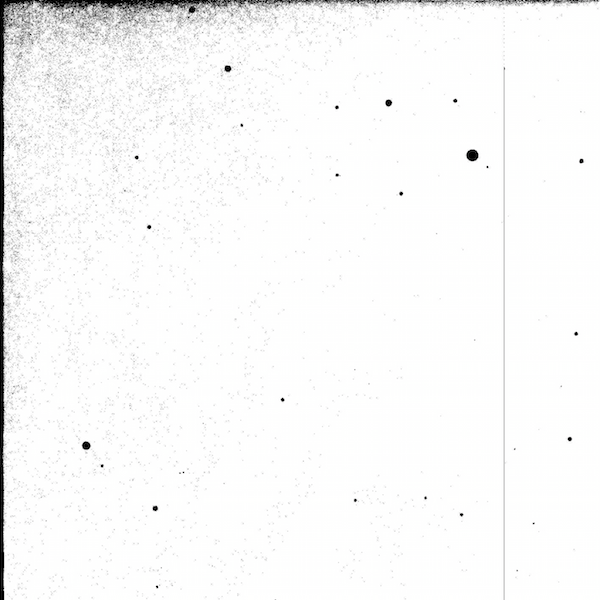}}\\
(a)\\
\fbox{\includegraphics[width=60mm]{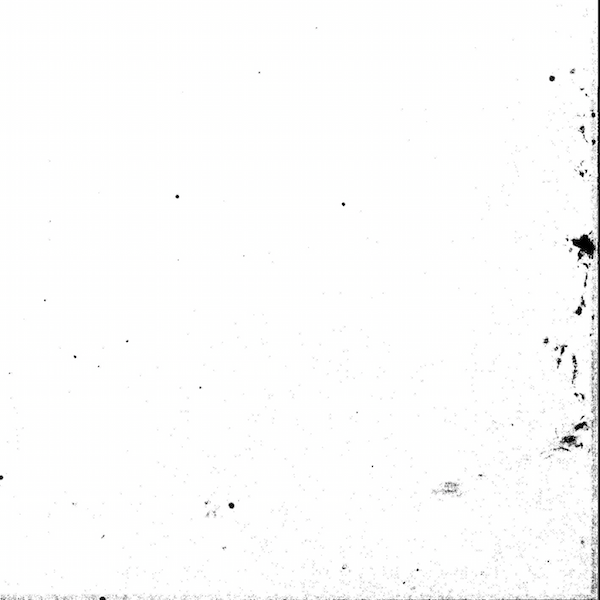}}\\
(b)\\
\fbox{\includegraphics[width=60mm]{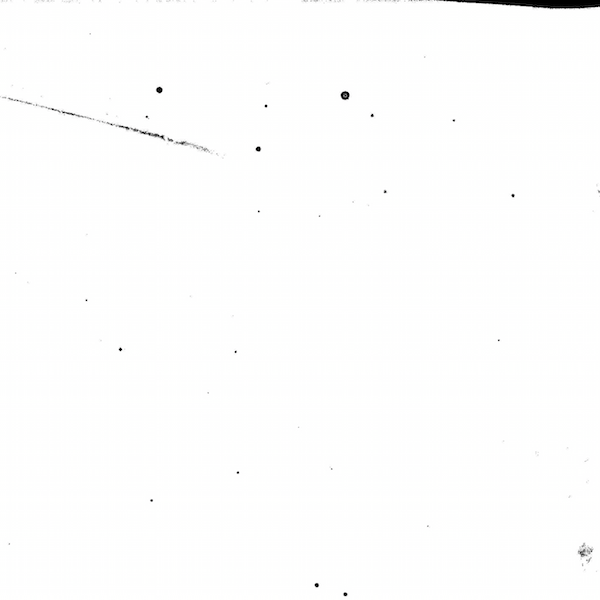}}\\
(c)
\end{tabular}
\caption{\label{fig:Mask}
A combined image was produced by taking median of 50 random images. This image was used as a mask, with zero-weights given to the spots during co-adding. The sections shown here are approximately a tenth of the original images.
[Example of mask produced during the process - its a binary image, with black pixels being masked away]
}
\end{figure}

Apart from cosmic rays, satellite trails, etc., an important artifact in images were spots due to dirt and dust. These regions consistently showed lower counts than neighboring pixels (for example, see Fig.~\ref{fig:CR}, bottom panel). These spots were present at the same pixel location over long periods. As the usual imaging procedure involved small dithers between exposures, we can address these problems by creating a bad pixel mask and excluding those from the final co-add. We created such a mask by co-adding 50 exposures of randomly selected fields (Fig.~\ref{fig:Mask}). We converted this into a binary mask by applying a simple sigma-clipping algorithm: all pixels in the mask with values 2$\sigma$ below the mean were marked as ``bad'' and assigned a boolean value 0, while all other pixels were assigned the value 1. This was used as the weight mask in the final SwarpA co-addition process. Owing to some individual exposures having 4 extra pixel columns, the mask --- a superset of all input images --- was larger than typical images to be processed. This issue was tackled by trimming two columns from each side of the mask before using it during co-addition.

To conclude, the four methods tested were similar in many respects, but we found that robust artifact removal was crucial for producing reliable co-added images. This could have been achieved by source code modifications in Montage, but was readily available in SWarp. After analyzing various parameters in the test sets --- e.g. time taken to co-add, modularity of the method, quality of the co-adds, we decided to use SwarpA to co-add the whole set of images.

 \subsection{\label{sec:com_sur}Comparing with other surveys}
\begin{figure*}
\centering
\begin{tabular}{ccc}
{\includegraphics[width=45mm]{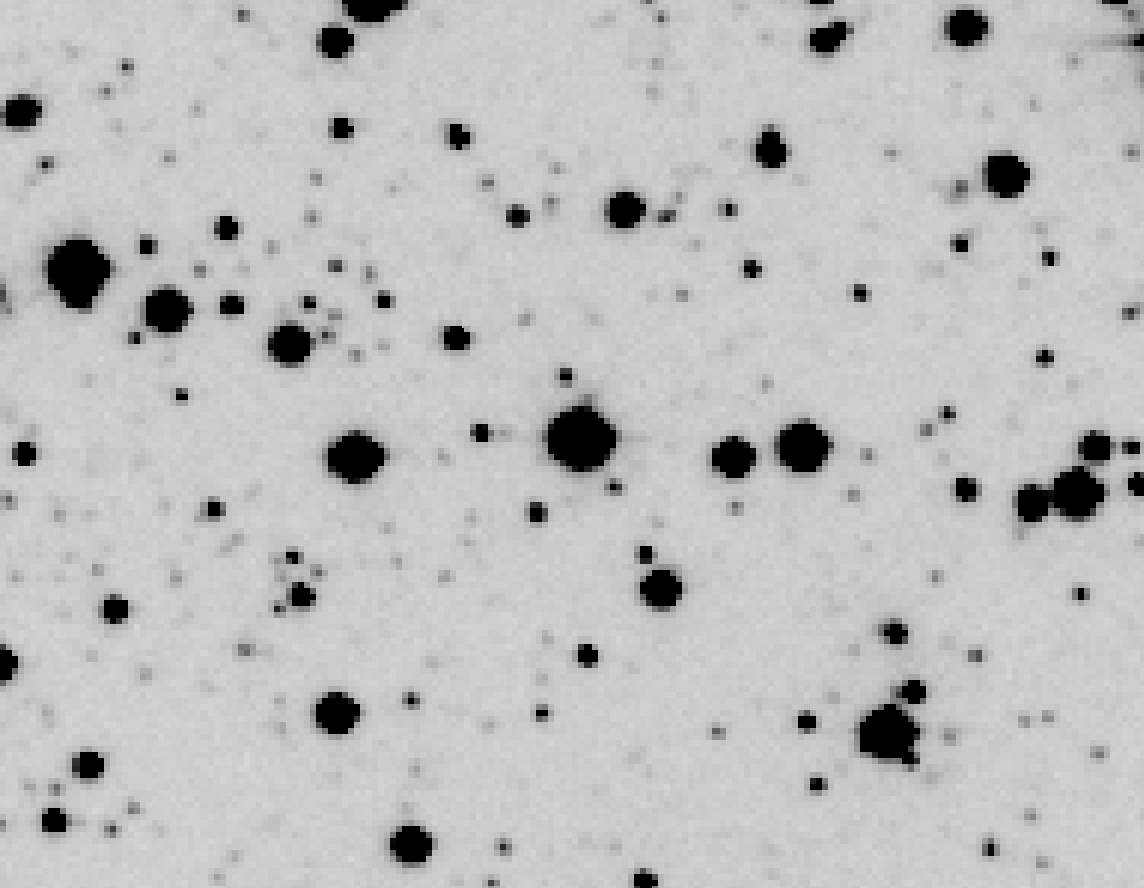}} & {\includegraphics[width=45mm]{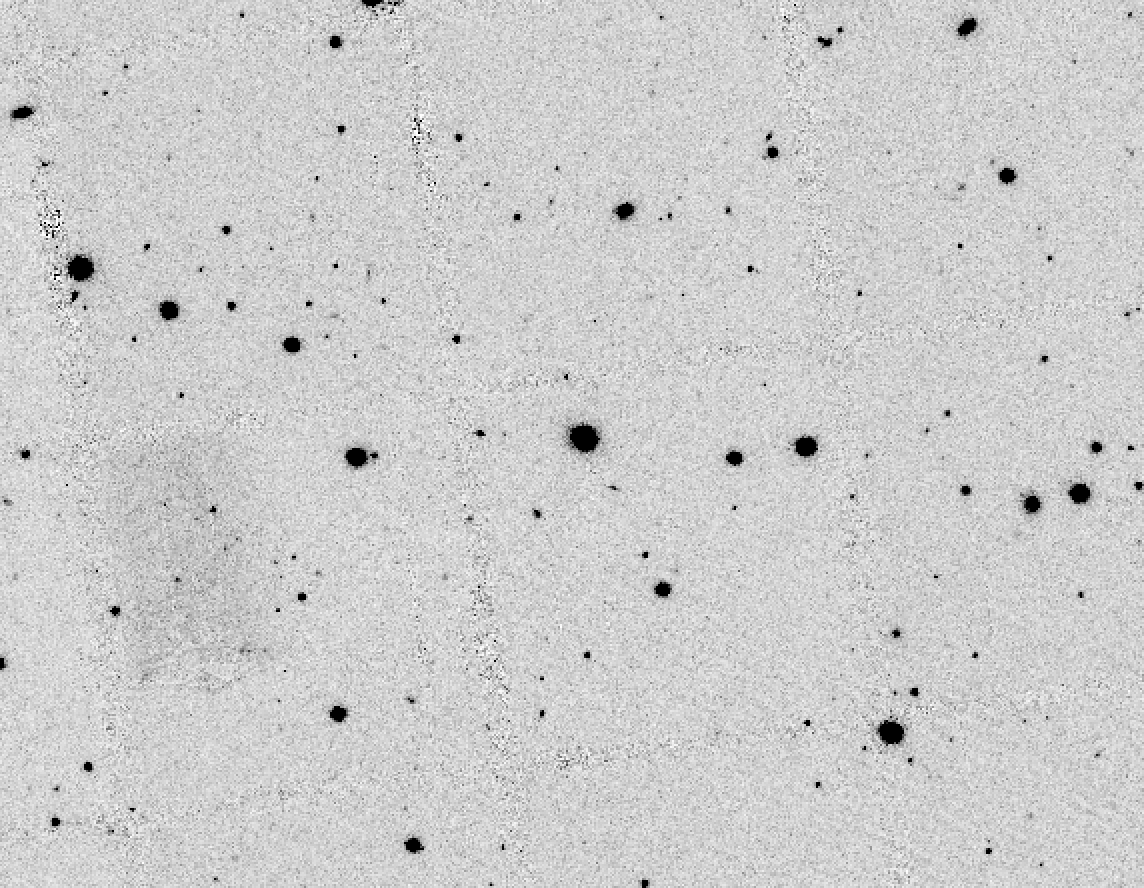}} & {\includegraphics[width=45mm]{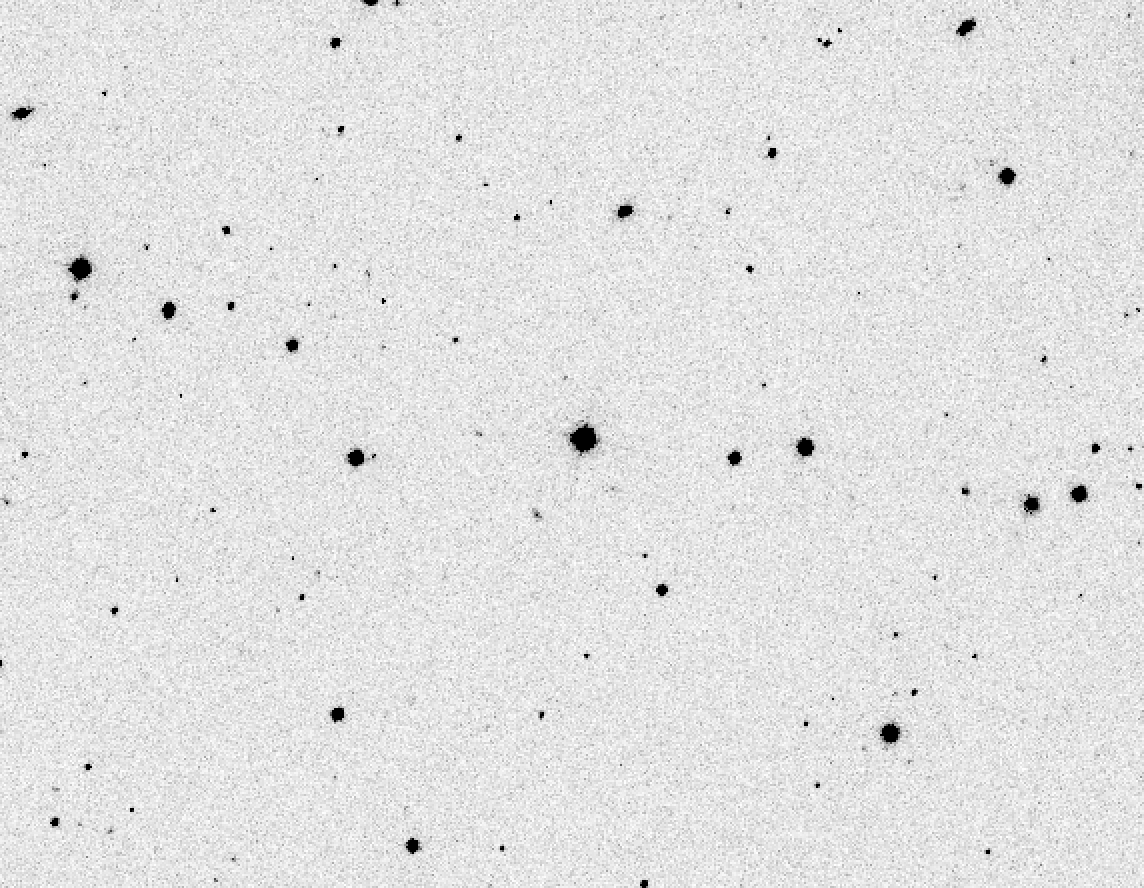}}\\
(a) & (b) & (c)
\end{tabular}
\caption{\label{fig:survey}
A small $\sim3\arcmin.8 \times 3\arcmin.0 $ postage stamp around $\alpha = 8^h 41^m 23^s, \delta = +17\degr 55\arcmin 33\arcsec$ (130.34622\degr, 17.926073\degr) with intensities z-scaled to compare three different sky surveys.
From left to right: a) co-added image from CRTS, b) corresponding cutout from PS1 in $g$-band, and c) SDSS survey image in $g$-band. The limiting magnitude comparisons were done on a larger area centred on these coordinates (Sec.~\ref{sec:com_sur}).
}
\end{figure*}

Once the preferred co-addition method is identified, we compared the CRTS co-added images with other all sky surveys\footnote{ZTF data were not publicly available when this study was started. }, such as Pan-STARRS1 (PS1) and SDSS. We arbitrarily  choose a region where all 3 surveys have an image with a reasonable overlap with each other. We arbitrarily picked the images to roughly centre around RA: 8h41m23s  Dec: +17d54m21s. Fig.~\ref{fig:survey}
shows (top to bottom) a cutout of a co-added image from CRTS, a cutout from the PS1 images and SDSS survey image respectively.

\begin{figure*}
\includegraphics[width=\textwidth]{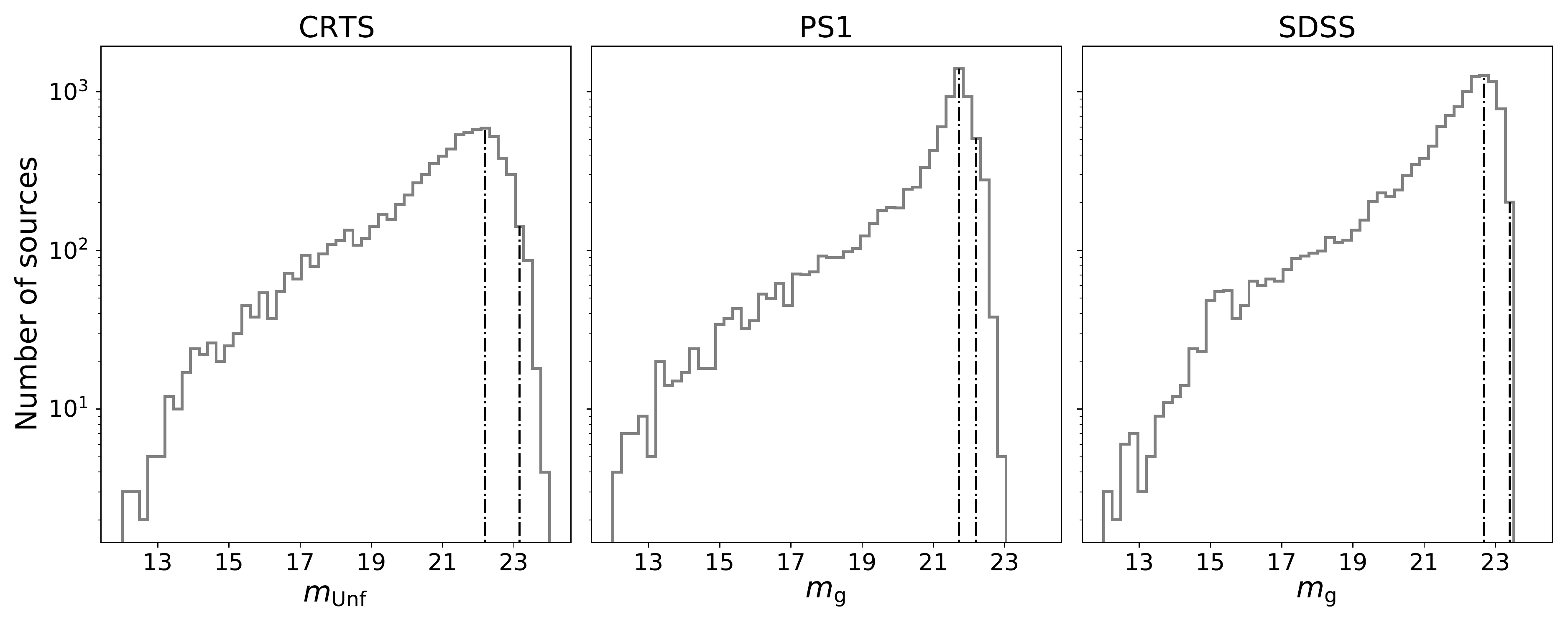}
\caption{Histograms of magnitudes of sources detected in various surveys in a 30\arcsec\ region centred at $\alpha = 8^h 41^m 23^s, \delta = +17\degr 55\arcmin 33\arcsec$ (Sec.~\ref{sec:com_sur}). The vertical dash-dotted lines show the mode and the halfway point as discussed in Fig~\ref{fig:mag_hist}. From left to right: unfiltered zero corrected CRTS, $g$-band sources in PanSTARRS, and $g$-band sources with $\Delta m_g < 0.2$ in SDSS. Number of sources within the histogram for CRTS, PS1, SDSS are 7774, 8004, 11852 respectively.}
\label{fig:survey_hist} 
\end{figure*}

For limiting magnitude comparison, similar to what is being described in Sec.~\ref{subsec:compare}, we extracted the CRTS sources and their magnitudes using the SExtractor and for PS1\footnote{{https://catalogs.mast.stsci.edu/panstarrs/}} and SDSS\footnote{{http://skyserver.sdss.org/dr16/en/home.aspx}}, catalog cone searches were used, within $30\arcmin$ around the mentioned RA, Dec 
in $g$-band.  
We see that CRTS is deeper than PS1 \cite{chambers2016pan}, where the stacked images have a limiting magnitude of 23.3. Comparison with SDSS \cite{lundgren2015sdss} is trickier: the limiting magnitude of SDSS DR16\footnote{\url{https://www.sdss.org/dr16/imaging/other_info/}} is 23.13, but the database includes sources detected with very low significance. To approximate the 5-sigma sensitivity, we have created a histogram only of those sources with a g-band photometric error $\Delta m_g < 0.2$.  In order to compare the the magnitudes of all 3 surveys, we would need to compute the zero point constant to the the SExtractor MAG\_ISO value computed for the CRTS sources, by cross-matching the common sources as shown in the Fig.~\ref{fig:survey_zero}. Fig.~\ref{fig:survey_hist} compares the histogram of limited magnitude for CRTS, PS1 and SDSS respectively of the sources within the region around the images in Fig~\ref{fig:survey}.

We note that CRTS images are unfiltered, with photometry corresponding loosely to Johnson's V-band - clearly distinct from $g$-band PS1 / SDSS magnitudes for our comparisons. We take this step instead of calculating filter transformations to match the CRTS band, as these comparisons are only meant to be indicative of the overall depth and sensitivity of CRTS images.

From the number of sources and limiting magnitudes of images in this analysis, we can conclude that CRTS co-add images can reveal fainter sources than PS1 and SDSS in many parts of the sky. However as Fig.~\ref{fig:survey} shows, the PSF of CRTS co-adds is a factor 2 wider than the others: with the average FWHM being 3\arcsec.1 (computed from SExtractor), as compared to  1\arcsec.39 for PS1 \citep{magnier2016pan}, and 1\arcsec.32 for SDSS \citep{ross2011ameliorating} respectively. This trade-off between deeper and sharper images will have to be resolved by the end user based on their requirements in image subtractions and transient searches.

Due to the differences in aperture size, total exposure, filter used etc. the total number of sources is often not the best metric when comparing a survey with a deeper survey. However, it is still important to do such a comparison if only to understand the fraction of sources that may be lost to noise, or blending, and to understand the usefulness of a survey for detecting, say, transients. We do so in the next section.

\subsection{Comparison with a deeper catalog}
Another way to investigate the sensitivity of our co-added images and catalog is to compare it with a significantly deeper catalog. This allows unambiguous discrimination between true and spurious sources, as well as provides a high quality photometric reference. For this purpose, we pick one of the deep fields in the CHFT Legacy Survey \citep{2012AJ....143...38G}. This survey includes ultra deep images of the Cosmos field, obtained with the 1 \sqd\ MegaCam. The catalog created from these CFHT images reaches 50\% completeness at $m_g = 27.9$ (AB magnitude): significantly deeper than our typical images.

This CFHT ``D2'' field, centred at $\alpha = 10^h 00^m 28^s, \delta = 02\degr 12\arcmin 30\arcsec$ overlaps with the N01054 CRTS field which was created by co-adding 382 survey images. We pick a 20\arcmin $\times$ 20\arcmin\ overlapping region from these fields for our analysis. To avoid problems with spurious sources near bright stars, we excluded a 3\arcmin.3\ region around every star brighter than $m_g = 14.5$. To get robust cross-matching, we picked a match radius of 4\arcsec: larger than diagonal of our 2\arcsec.5~pixels. We follow a method similar to Sec.~\ref{sec:com_sur} and derive a zero-point offset of 27.3~mag.

\begin{figure}
\includegraphics[width=8cm]{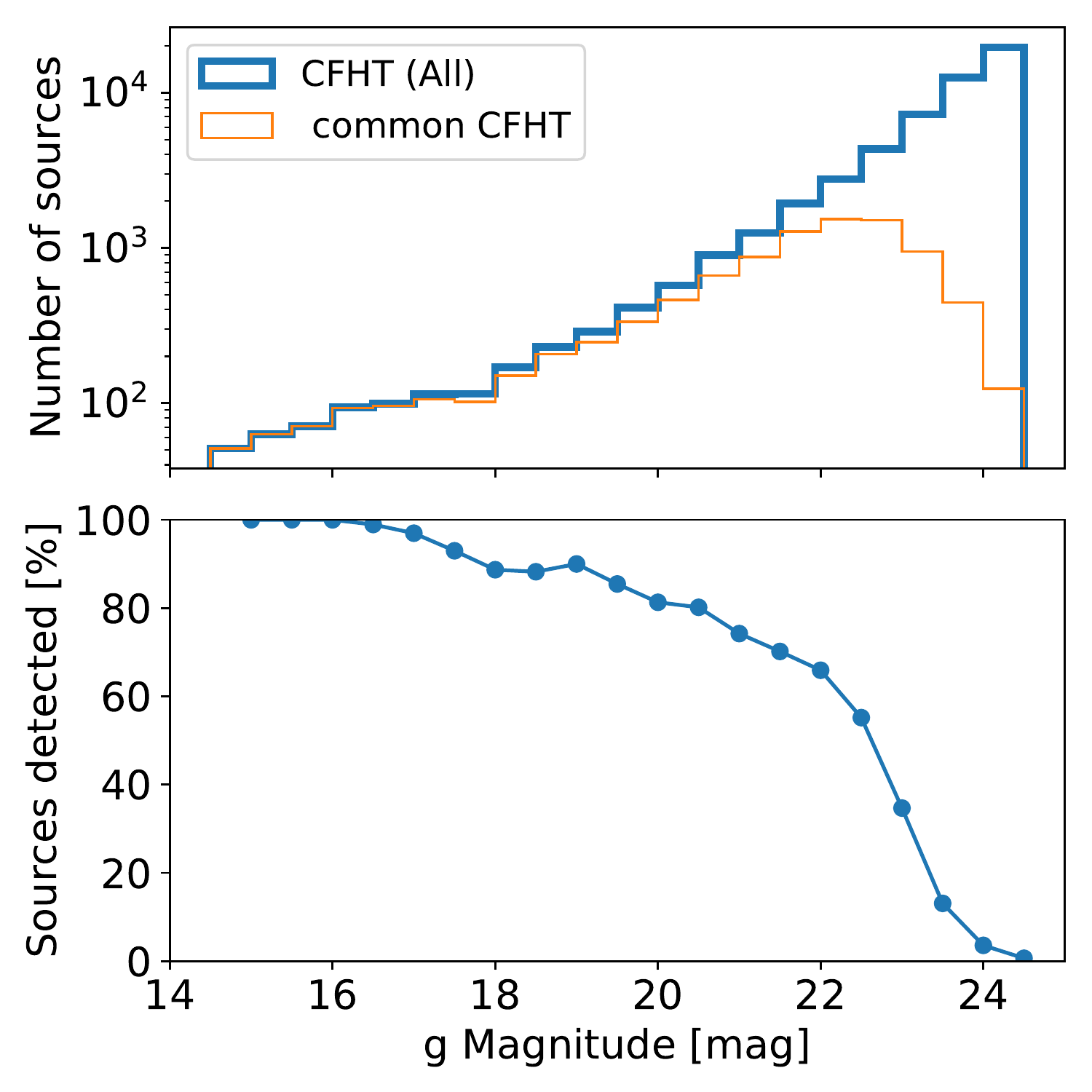}
\caption{\textit{Top:} Comparing the complete CFHT data set for the Cosmos field (blue line) with the subset of stars that were also found in our CRTS catalog (orange line). CFHT data are truncated at 24.5~mag, as there are no CRTS sources fainter than this value. \textit{Bottom:} The fraction of CFHT sources that were detected in CRTS, calculated in 0.25~mag bins. We reach 50\% coverage at about 22.6 mag for this field, created by co-adding 382 images.}
\label{fig:det} 
\end{figure}

Since the CRTS images are much shallower than the CFHT catalog, we assume that any source present in CFHT is a true source. First we find the number CFHT sources that are present in our CRTS catalog. The top panel of Fig.~\ref{fig:det} compares the total number of sources found in the CFHT catalog (blue line) with the subset of these sources that was also detected in our catalog (orange line). We see that the catalog is nearly complete till $\sim 22$ magnitude. This completeness is better demonstrated in the lower panel, which shows the percentage of CFHT sources recovered in CRTS images. Our image of this field has 50\% completeness at about 22.6~mag. We note that this depth varies over the sky based on the number of images that were co-added.

Having checked the completeness of the catalog, we now turn our attention to validity.
The upper panel Fig.~\ref{fig:det} compares the histogram of all sources in the CRTS catalog (blue line) with those that are also present in the CHFT catalog (orange line) - and the curves nearly overlap. The lower panel quantifies any discrepancies in terms of a false alarm fraction:  we see that this fraction remains at a few percent level across all magnitudes.  A plausible explanation of poor coverage at some magnitude might be poor PSF, which blends in the nearby objects to the SExtractor algorithm.

\begin{figure}
\includegraphics[width=8cm]{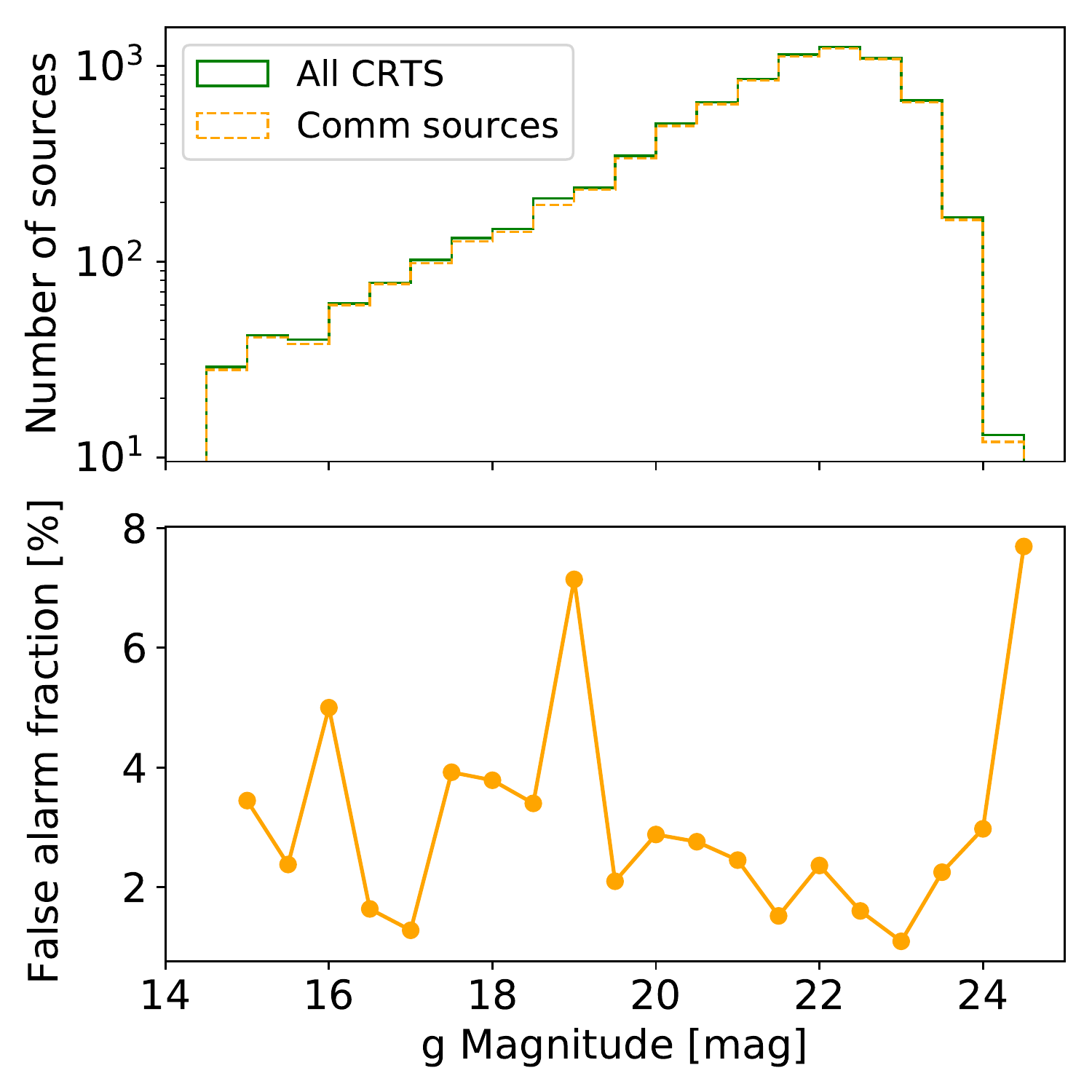}
\caption{Evaluating the validity of the CRTS catalog. \textit{Top:} The green line shows a histogram of all sources found in the CRTS image, and the dashed orange line shows those CRTS sources that are also present in the CFHT catalog. We see a good overlap between the two. \textit{Bottom:} The fraction of sources detected in the CRTS image which were not present in the CFHT image, evaluated in 0.5~mag bins. We see that the false positive rate is at a few percent level, independent of magnitude.}
\label{fig:far} 
\end{figure}

\section{Implementation}\label{sec:implementation}
We now discuss the co-addition process in detail, including the workflow (\S\ref{subsec:workflow}), hardware used for processing (\S\ref{subsec:hardware}), processing time (\S\ref{subsec:performance}) and accessing the final products (\S\ref{subsec:finalstack}).

\begin{figure}
\centering
\includegraphics[width=\columnwidth]{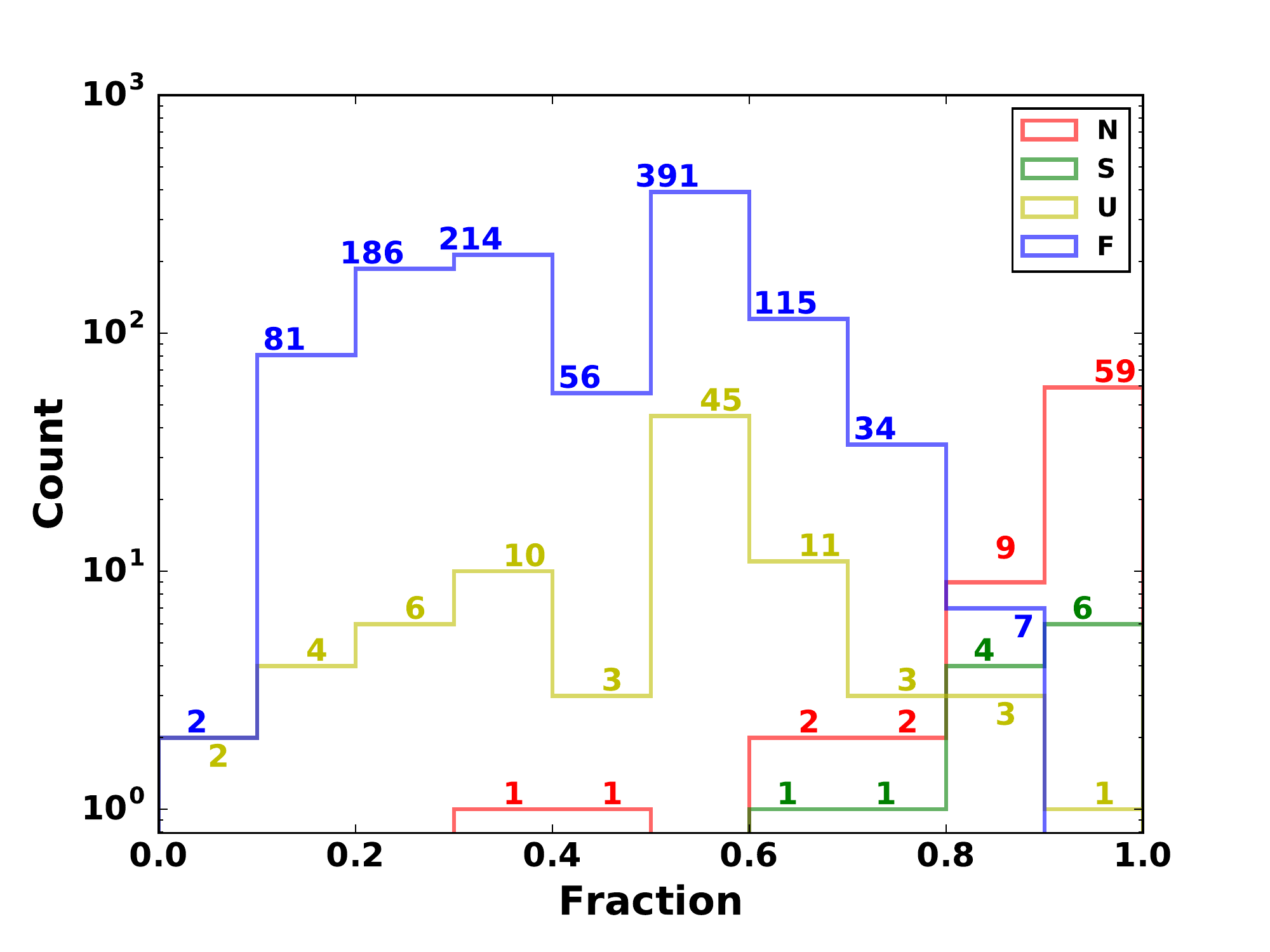}
\caption{\label{fig:scaa}
Of the total 7894 fields, 1260 fields were identified to have a high scatter in pointing directions across individual images. The co-adds for these fields were created by excluding the outlying fields from the final stack. The histograms here show the number of fields as a function of the fraction of total exposures retained in the final stack. For instance, 59 N fields with high scatter retained $>90\%$ images in the final stack, 9 fields retained 80--90\%, and so on. $U$ and $F$ fields followed moving objects resulting in larger scatter and fewer of those fields got included in the final co-adds. Co-added images resulting from individual images with large angular separation result in large contiguous rectangular regions containing both the images. The process is computationally resource hungry with only an incremental advantage and hence such outliers were ignored.  }
\end{figure}

\subsection{Workflow}\label{subsec:workflow}
The setup at IUCAA is such that images are stored at one place, and were to be processed at another. Hence, the first step was to copy all images to a staging area on the source machine, \texttt{scp} them to the co-add server into a folder for a given field, then clear the staging area. We H-decompressed the images,
verified the presence of a valid WCS in them, and ensured pixel dimensions as indicated in the header were equal to that of the image mask.  In the few cases where these conditions were not met, the images were excluded from the final co-add. A text file containing names of all valid images was generated for each field, and SWarp configuration file was also generated. Most parameters were left at default values, with notable exceptions listed in Table~\ref{tab:listing}.
The master script then calls SWarp for co-addition.

\tikzstyle{decision} = [diamond, draw,
    text width=4.5em, text badly centered, node distance=3cm, inner sep=0pt,fill=red!20]
\tikzstyle{block} = [rounded rectangle, draw, fill=blue!20, 
    text width=5em, text centered, minimum height=1.5cm,text width=2cm]
\tikzstyle{line} = [draw, -latex']
\tikzstyle{cloud} = [draw, ellipse,fill=red!20, node distance=3cm,
    minimum height=1cm,  text width=2cm,text centered]
    \tikzstyle{cloud2} = [draw, ellipse,fill=orange!30, node distance=3cm,
    minimum height=1cm,  text width=2cm,text centered]
\tikzstyle{io} = [trapezium, trapezium left angle=70, trapezium right angle=110, minimum width=1cm, minimum height=1cm, text centered, draw=black, fill=green!30,  text width=1.5cm]
\tikzstyle{process} = [rectangle, minimum width=2cm, minimum height=1cm, text centered, draw=black, fill=orange!30,  text width=2cm]
\tikzstyle{process2} = [rectangle, minimum width=2cm, minimum height=1cm, text centered, draw=black, fill=red!20,  text width=2cm]
 \tikzstyle{cloud3} = [draw, circle,fill=blue!20, node distance=3cm,
    minimum height=1cm,  text width=2cm,text centered]
\tikzstyle{man} = [trapezium, trapezium left angle=70, trapezium right angle=70, minimum width=1cm, minimum height=1cm, text centered, draw=black,  fill=blue!20,  text width=2cm]

\begin{figure*}
  \centering
\begin{tikzpicture}[node distance = 2cm, auto]
    \node [process] (scp) {SCP copied images to cluster};
    \node [block, left of=scp, node distance=3.5cm, fill=orange!30] (ssh) {SSH; to copy images on home directory};
    \node [process2, right of=scp, node distance=3cm] (rm) {Remove images from home directory};
    \node [process, below of=scp,fill=blue!20,
    minimum height=1.5cm] (hd) {hDecompress images};
    
    \node [block, left of=hd, node distance=3.5cm] (wcs) {Discard invalid WCS images};
    \node [block, left of=wcs,node distance=3.5cm] (mask) {Discard if does not match the Mask dimension};
    
    \node [io, below of=wcs, node distance=2.5cm] (inp) {Generate list of images to co-add};
    \node [io, right of=inp, node distance=3.5cm] (clust) {Select images closely scattered};
     \node [man, left of=inp, node distance=3.5cm] (con) {Configuration and Mask file};
     \node [process, below of=inp, node distance=2.5cm,fill=red!50] (co-add) {co-add using SwarpA};
      \node[decision, below of=clust, node distance=2.5cm](dec){is size $\geq$ 700MB};
      \node [process2, right of=dec, node distance=3cm] (rm2) {remove all temp files};
      \node [block, right of=rm2, node distance=3.5cm, fill=orange!30] (dd) {Dump co-adds into external Disk};
    \path [line] (ssh) -- (scp);
     \path [line] (scp) -- (rm);
     \path [line] (rm) |- (hd);
     \path [line] (hd) --  (wcs);
     \path [line] (wcs) -- (mask);
     \path [line] (mask) --(inp);
     \path [line] (clust) -- (inp);
     \path [line] (con) -- (inp);
     \path [line] (inp) -- (co-add);
     \path [line] (co-add) -- (dec);
     \path [line] (dec) -- node[anchor=west]{Yes} (clust);
     \path [line] (dec) -- node[anchor=north]{No} (rm2);
      \path [line] (rm2) -- (dd);
\end{tikzpicture}
\caption{Flow chart of the work flow used to execute the co-addition. Colors of the nodes represent the following. Orange represents, steps involving remote computers, \textit{i.e.}, not part of the cluster where co-addition takes place. Pink represents steps carried out only due to memory constrains, hence Pink and Orange processes are not mandatory and can be avoided if performing in a local directory. Violet and Green are mandatory steps, with the difference being Green steps may require correction. And lastly Red presents the process where core algorithm takes place.
}
\end{figure*}
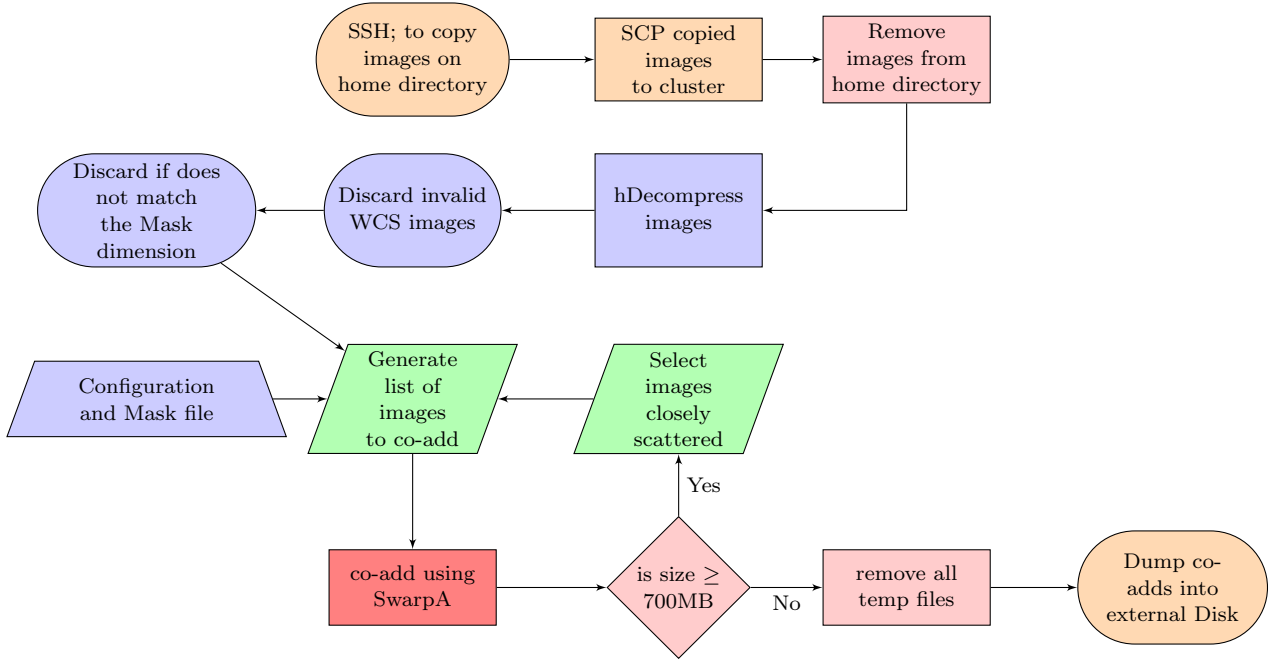

\begin {table*}
\label{tab:listing} 
\begin{center}
\begin{tabular}{llp{7cm}}
\hline
Keyword & Value & Notes \\
\hline
WEIGHT\_THRESH & 0.5 & Mask pixels with value $<0.5$ will be ignored in the co-add. The actual mask is binary, with zeros for bad pixels and ones for good pixels\\
COMBINE\_TYPE & CLIPPED & Use the clipping algorithm as per \citet{clip}\\
CLIP\_AMPFRAC &0.3 & Default value recommended by \citet{clip} \\
CLIP\_SIGMA & 4.0 & Default value recommended by \citet{clip}\\
SUBTRACT\_BACK & Y & Removes background, eliminating most background-related artifacts from co-added image\\
DELETE\_TMPFILES & Y & Deletes the temporary files created during the co-add process.\\
IMAGEOUT\_NAME & [Field\_ID]\_clip\_mask\_SBG\_Y.fits &  The suffix signifies the major algorithm properties: sigma clipping, masking input images, and background subtraction\\
WEIGHTOUT\_NAME  & [Field\_ID]\_clip\_mask\_SBG\_Y.weight.fits &\\
WEIGHT\_IMAGE & Mask\_S.fits & The name of the mask file shown in Fig. \ref{fig:Mask}\\
\hline
\end{tabular}
\end{center}
\caption{Key parameters for SwarpB configurations
}
\end{table*}

\begin{figure}
\includegraphics[width=\columnwidth]{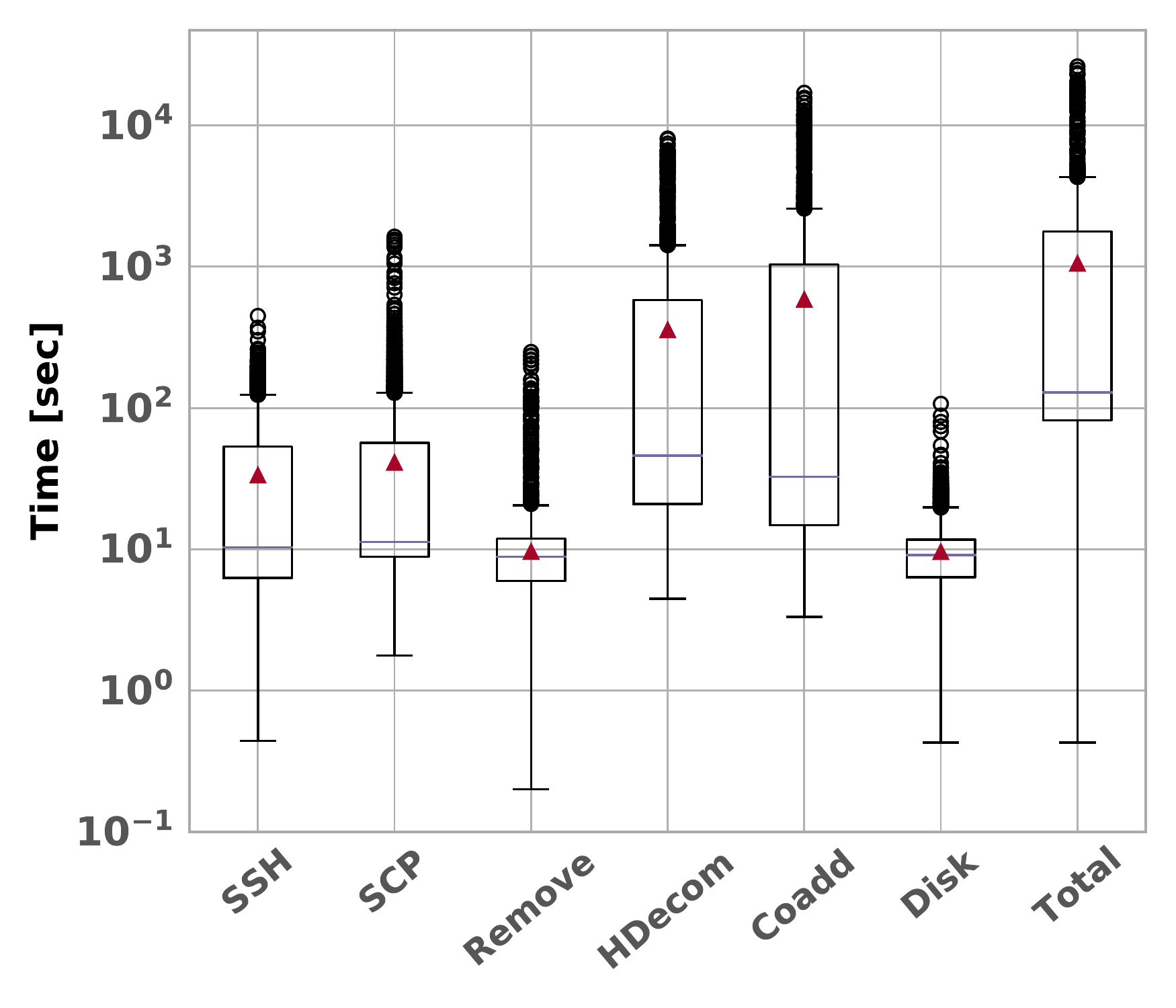}
\caption{Box plot of time taken (in seconds) by each of the co-addition stages for different fields. In some cases the contributors to the dispersion are mainly the number of images in a field (and the resulting sizes, operations etc.), but another factor is also whatever else the machine may be handling at the time. For each column, the red line and the red square represent median and mean values respectively. The height of the box indicates the inter-quartile range (first to third quartile), and the outliers are shown by the `o' markers. In general, the data decompression took the same amount of time as the co-addition itself, and typically over an order of magnitude more than the other tasks. }
\label{fig:box} 
\end{figure}

SWarp then reads all input images along with their weight maps. In our case, these weight maps are simply binary masks intended to eliminate known CCD artifacts. SWarp uses this information to build a background map and subtract it from the images. All input images are then re-sampled and projected onto an appropriate subsection of the output frame rejecting the masked pixels. The images were co-added with sigma--clipping mean to reject outliers. Since all input images had the same weight, no weight files were used for this step. However, if hierarchical co-addition is performed (for instance \S\ref{subsec:montage}), then appropriate weight files generated in each intermediate step should be supplied to the consecutive steps. 

The master script deletes temporary files generated by SWarp, retaining only the final co-add and the corresponding weight file. The master script logs all operations for debugging and statistical purposes. It also adds  appropriate \texttt{FITS} header keywords to the final co-added image, describing the processing involved (Table~\ref{tab:headers}).

Typical output images are 70-100 megabytes in size, as the area varies a little. In some cases, we found that the final image was extremely large -- this happened if one of the input images had a large on-sky offset from the typical images in the input stack. Such an offset may arise either from a wrong (but semantically valid) WCS in the image, or by incorrect labeling of the field ID for an input image. As a simple cut for identifying such images, we flagged cases where the co-add was larger than 150 MB. This was noticed for  450 fields out of 7894 (5.7\%). Over time we will remove them using visual inspection, a time consuming task deferred for the future. Another future task is to exclude from co-addition images with poor seeing and/or high background sky. This is expected to marginally reduce the number of blends seen in the current co-adds.

\begin{table*}
\centering
\begin{tabular}{l|l}
\hline
Keyword & Value\\
\hline
AUTHOR & akshat.singhal014@gmail.com \\
COADDN & number of input images (number of images co-added)\\
DATET & Date and Time of creation of file \\
INSTI & `IUCAA' \\
EPOCH & 2000 \\
DETECTOR & `IMAGER LABS' \\
FILTER & `NONE' \\
TELESCOP & `CATALINA SCHMIDT 003' \\
OBSERVAT & `UA Bigelow Station' \\
CRTSDATA & Total number of images in CRTS repository for this field \\
\hline
\end{tabular}
\caption{Keywords added in the final CSS image co-adds.\label{tab:headers}}
\end{table*}

\subsection{Hardware / Computing infrastructure}\label{subsec:hardware}

The CSS images are processed into catalogs by Lunar and Planetary Laboratory (LPL)  in Tucson, Arizona to look for NEOs and then passed on to Caltech to look for transients. At Caltech we obtain 5\arcmin $\times$ 5\arcmin\ cutouts centered around the found transients. The full images of the CSS are transferred to the Inter-University Center for Astronomy and Astrophysics (IUCAA), in Pune, India and served in the form of 5\arcmin $\times$ 5\arcmin\  cutouts\footnote{http://crts.iucaa.in}  through a web interface which uses a Python/Flask backend. The web application serving these cut-outs is hosted on an HP Blade server with dedicated storage class server for the imaging data and a dedicated server for the photometric catalog and the imaging metadata. Originally we planned to use these cutouts for co-adding, but the small dithering between revisits meant that each cutout would have non-uniform edges and varying SNR.  Initial experiments also revealed that that process would take up to 50 times longer as a result of the larger number of images.

We used `Perseus', a 64-node cluster at the High Performance Computing (HPC) Center at IUCAA, for our stacking experiments. Each node of Perseus has two CPUs with eight cores each, each node having 128GB RAM. Ideally it should have been possible to execute all co-addition jobs in parallel as the number of cores, 1024, is of the same order as the number of fields viz. 7894. However, moving the images to Perseus in real-time turned out to be a bottleneck. We discovered that at most 30 processes could be run concurrently. A dedicated storage could have been attached to Perseus but it was not possible at the time of this work as the cluster was also being shared by other processing jobs. In the end we typically used 20--25 cores at a time for the co--additions. The total time required was $\mathcal{O}$(2 weeks).

\subsection{Performance / time taken}\label{subsec:performance}

We now discuss the time taken by different processes. The times taken to SSH, SCP the images and transferring the co-adds to the disk were the fastest compared to H-decompress and co-addition processes. Fig.~\ref{fig:box} shows a box plot of time taken per co-add by the individual processes and total time from remotely accessing images to transferring co-adds to an external hard disk.

Once we settled on the method and the set-up and debugging was done, the total time for the co-adds was 10 clock-days. Of the 7894 fields we started with, 7791 fields were successfully co-added.

\subsection{Final stack}\label{subsec:finalstack}

\begin{figure*}
\centering
\begin{tabular}{c}
\includegraphics[trim=0 120 0 120,clip, width=\textwidth]{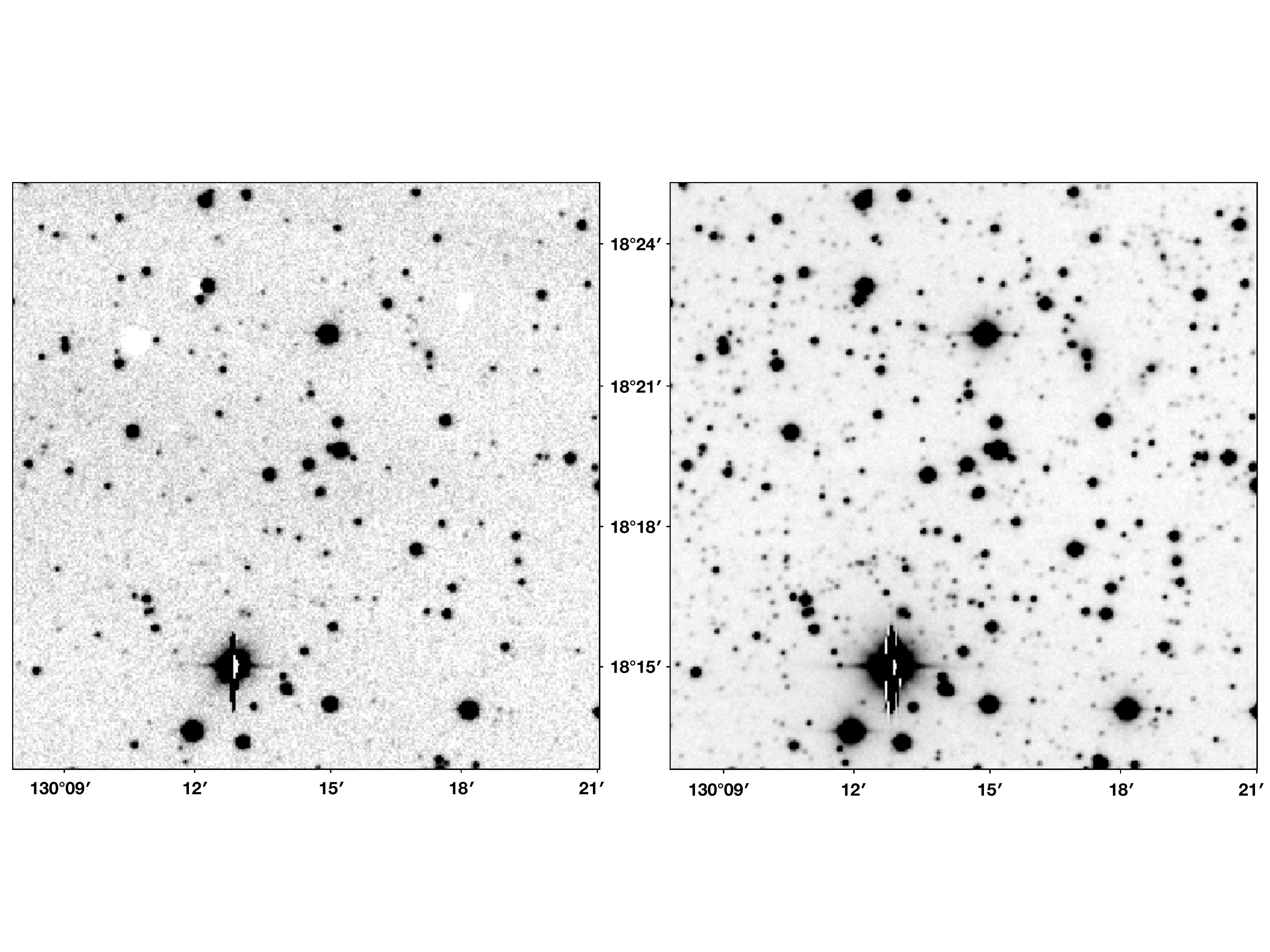}\\
\includegraphics[trim=0 120 0 120,clip, width=\textwidth]{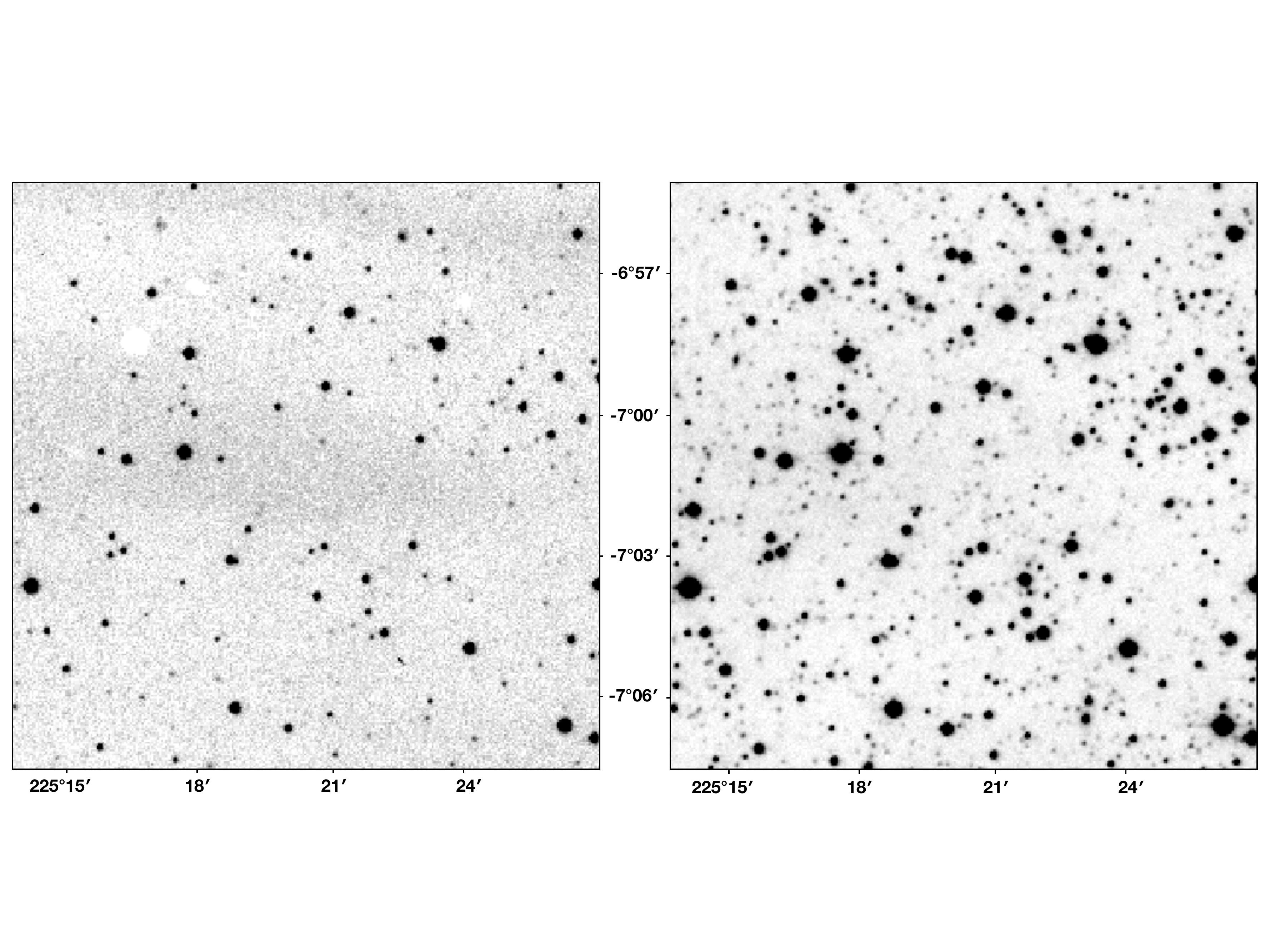}
\end{tabular}
\caption{\textit{Top row :} A section of field N18045 $\sim12'$ on a side around RA, Dec 08h40m57.9354s +18d19m02.4383s. \textit{Top left :} you see a single-epoch image, and \textit{Top right :} you see the same section for the co-added image. This field had only 439 images, out of which 497 have gone in to the co-added image. \textit{Bottom row :} A section of field S07080 $\sim12'$ on a side around RA, Dec 15h01m21.1668s -07d01m17.2091s. \textit{Bottom left :} you see a single-epoch image, and \textit{Bottom right :} you see the same section for the co-added image. This field had only 268 images, out of which 299 have gone in to the co-added image. Note that defects in the single image are not seen in the co-added image.}
\label{fig:co-adds}
\end{figure*}

The total size of the co-added images is 1.5~TB, and the images cover an area of $\sim$27000 sq. degrees. Users can access the images from \url{http://crts.iucaa.in} . Users who find these data useful in their research are requested to acknowledge it as ``\textit{This research makes use of data from the Catalina Sky Surveys, LPL, UA and Catalina Realtime Transient Survey, Caltech, with value additions and served by IUCAA through a web application currently hosted at \url{http://crts.iucaa.in}.}'' Fig.~\ref{fig:co-adds} shows the comparison of single image vs stacked methodology as discussed.

\section{Concluding Comments}\label{sec:future}
We have carried out the co-addition of CSS images to produce very deep co-adds with up to $\sim 3$ magnitudes more depth than a single image. CRTS also includes the Mount Lemmon Survey (MLS) and Siding Spring Survey (SSS) surveys. We plan to co-add these images as well to provide coverage in the South, and greater depth. CSS and MLS cameras are currently using larger, 10k x 10k chips. In the near future we will be co-adding images from these cameras as well.

A comparison with additional combining softwares will provide an insight about what should be used to continually improve reference images over large areas in the era of Zwicky Transient Facility \citep[ZTF;][]{bellm14} and Large Synoptic Survey Telescope \citep[LSST;][]{LSSTSciBook2009}.

The cutouts will be served along with the images for individual epochs already being served from the CRTS public server at IUCAA\footnote{http://crts.iucaa.in}. The set comprising the deep image in its entirety is 1.5~TB.

In order to realize the full potential of the stacked images, a catalog needs to be obtained. Achieving this however is non-trivial since the depth of the co-adds varies from field to field. Until such a catalog is produced, the users will have to rely on local calibration, or use the deep images for detection purposes.

\section*{Acknowledgements}

We thank, Bruce Berriman, John Good, Frank Masci, Steve Hartung and Simon Krugoff for their guidance.  We also thank CSS, CRTS, IUCAA, IUSSTF and NSF for the excellent support.

The Catalina Sky Survey is funded by a grant from the National Aeronautics and Space Administration’s Near-Earth Object Observations program. The CRTS web service at IUCAA was partly developed under the \textit{Virtual Observatory} program funded by the Ministry of Communication and Information Technology of the Government of India and the \textit{Data Driven Initiatives} program funded by the National Knowledge Network.

This work is based in part on data products produced at Terapix available at the Canadian Astronomy Data Centre as part of the Canada-France-Hawaii Telescope Legacy Survey, a collaborative project of NRC and CNRS.

AAM, AJD, CD, MJG, and SGD were supported in part by the NSF grants AST-0909182, AST-1313422, AST-1413600, and AST-1518308, and the Ajax Foundation. AAM, MJG, and SGD were also supported by IUSSTF. We also thank ClassACT: Indo-US centre for astronomical object and feature characterization and classification.

This research made use of Montage. It is funded by the National Science Foundation under Grant Number ACI-1440620, and was previously funded by the National Aeronautics and Space Administration's Earth Science Technology Office, Computation Technologies Project, under Cooperative Agreement Number NCC5-626 between NASA and the California Institute of Technology. 
This work was done during the tenure of Akshat Singhal and Varun Bhalerao in Inter-University Centre for Astronomy and Astrophysics.

\section*{Data Availability}
All the final co-add images and weight files are available at \url{http://crts.iucaa.in}. Additional material such as configuration files, scripts for processing bulk  images, log files etc. are available at \url{https://ddi.iucaa.in/static/crts/}. The raw data underlying this article will be shared on reasonable request to the corresponding author. 

\bibliographystyle{mnras}
\bibliography{main}

\end{document}